\documentclass[lettersize,journal]{IEEEtran}
\usepackage{amsmath,amsfonts}
\usepackage{array}
\usepackage[caption=false,font=normalsize,labelfont=sf,textfont=sf]{subfig}
\usepackage{textcomp}
\usepackage{stfloats}
\usepackage{url}
\usepackage{verbatim}
\usepackage{graphicx}
\usepackage{cite}
\usepackage{xcolor}
\usepackage{color}
\usepackage{color,soul}
\usepackage{algorithm}
\usepackage{algpseudocode}
\usepackage{float}
\usepackage{tikz}
\usepackage{pgfplots}
\usepackage{svg}
\pgfplotsset{compat=newest} 
\newcommand{\nj}[1]{{\color{black} #1}} 

\newtheorem{remark}{Remark}
\newtheorem{definition}{Definition}
\begin{document}
	
	\title{
		Diverse Safety-Critical Scenario Generation for Automated Vehicles:\\ 
		A Feature-Driven Optimization Approach}

\author{
	N. Javanmardi\textsuperscript{1,*},
	N. Rajesh\textsuperscript{1},
	K. Tom Kurian\textsuperscript{1},
	E. Lefeber\textsuperscript{1},
	J. Ploeg\textsuperscript{2},
	N. van de Wouw\textsuperscript{1},
	M. Alirezaei\textsuperscript{1,3}%
	\thanks{\textsuperscript{1}Eindhoven University of Technology, Dept. of Mechanical Engineering
		Dynamics and Control group, Eindhoven, The Netherlands.}%
	\thanks{\textsuperscript{2} 2getthere Netherlands B.V., Helmond, The Netherlands.}%
   \thanks{\textsuperscript{3} Roadscor Netherlands B.V., Helmond, The Netherlands.}%
		\thanks{*Corresponding author:\, n.javanmardi1@tue.nl.}%
		\thanks{This work was performed as part of the Synergies project, funded by the European Union’s Horizon Europe Research and Innovation Actions under grant agreement no. 101146542.}%
	
}
	
	
	
	\maketitle
	\begin{abstract}
		This paper develops an approach for the generation of diverse, safety-critical scenarios for scenario-based testing in automated vehicles. We propose a behavioral modeling framework to formalize an optimization problem that supports such scenario generation. Then, we exploit a clustering-based evolutionary algorithm to identify diverse local and global optima. In particular, we define the optimization problem with a criticality metric as the cost function. The solution set of this multimodal optimization problem, which includes multiple optima, represents unsafe scenarios in terms of the criticality measure. 
		In addition, we employ a feature-based scenario dissimilarity metric to generate a set of safety-critical scenarios that are also diverse.
		To this end, a species-based evolutionary algorithm is adapted for the formulated optimization problem, incorporating the feature-based dissimilarity metric into its search strategy.
		Besides, we evaluate the performance of the proposed method in exploring diverse, safety-critical scenarios with respect to the features using a diversity measure. Accordingly, we define a feature-based diversity measure based on the dissimilarity metric.
		The applicability of the proposed methods is demonstrated through a simulation-based use case.
	\end{abstract}

	\section{Introduction}
Automated Vehicles (AVs) promise to provide accessible transportation while enhancing comfort, efficiency, sustainability, and safety. To achieve widespread public acceptance, AVs must demonstrably outperform human drivers, particularly in terms of safety \cite{nees_safer_2019}.

Guaranteeing safety in AVs is a complex challenge that currently requires millions of kilometers of on-road testing \cite{kalra_driving_2016}.  In addition to on-road testing, several alternative approaches have been developed to assess AV safety, including formal verification, shadow testing, and scenario-based testing \cite{riedmaier_survey_2020}. Scenario-based testing enhances the safety verification and validation of AVs by generating challenging and critical driving scenarios that simulate system performance across the vehicle's Operational Design Domain (ODD) \cite{riedmaier_survey_2020}. By focusing on these safety-critical scenarios, the number of test cases is reduced, consequently minimizing testing time and cost \cite{klischat2019generating,klischat2020scenario,althoff2018automatic}.

Generating safety-critical scenarios for scenario-based testing in AVs has been extensively studied in the literature, resulting in a variety of approaches. \nj{ As categorized in \cite{ding2023survey}, existing approaches include data-driven generation  \cite{ding2020cmts,zofka2015data}, adversarial generation \cite{chen2021adversarial,koren2018adaptive}, and knowledge-based generation \cite{klischat2019generating,klischat2020scenario,althoff2018automatic}, and  \cite{menzel2018scenarios}. Across these categories, the objective of safety-critical scenario generation can be mathematically unified by an optimization formulation \cite{ding2023survey}. Compared with methods such as grid search, which exhaustively sample possible scenarios to identify safety-critical ones, optimization-based methods can efficiently generate a limited set of safety-critical scenarios.}
	
The optimization problem is typically formulated using a metric function that serves as a cost function to quantify risk levels and safety-related properties of scenarios \cite{ding2023survey}. Criticality metrics--- categorized into time-based, deceleration-based, and energy-based types \cite{westhofen2023criticality,wang2021review}--- can be employed as cost functions in  optimization problems for scenario-based testing to generate safety-critical scenarios across a vehicle’s ODD \cite{singh2024systematic}.

However, most optimization-based studies represent scenarios as static variables and formulate the corresponding optimization problems using algebraic or static constraints on these variables  \cite{ding2023survey}. Therefore, they do not capture the temporal evolution of the scenarios and the associated dynamical and operational system constraints, which may lead to  dynamically infeasible optimized scenarios.

\nj{To explicitly define an optimization problem and the notion of a driving scenario, within a unified mathematical framework, we utilize the framework described in \cite{kevin}, which builds on the behavioral modeling approach introduced in \cite{willems1997introduction}.  The behavioral modeling approach introduces a behavioral set to represent a real-world physical system through a set of trajectories that satisfy behavioral relations governing interactions among system elements. Accordingly, \cite{kevin} applies this behavioral modeling approach to automated driving systems, introducing formal terminology for scenario-based testing in traffic environments, such as concrete scenarios. Using these definitions, we mathematically formulate scenario generation as a multimodal optimization problem over a unified behavioral set, guaranteeing that all safety-critical scenarios are subject to vehicle dynamics, physical constraints, and operational requirements. We also define a scenario as a specific trajectory within this behavioral set.}
Besides, we represent the criticality metric evaluated over the entire trajectory, providing a more realistic assessment of system safety. Hence, we formulate the optimization problem with explicit time-dependent dynamics and system constraints \nj{within a behavioral set}, ensuring that the resulting optimized scenarios are physically realizable.

Due to the high complexity and dimensionality of real-world traffic environments, the distribution of safety-critical scenarios generally represents a multi-modal optimization problem, in which multiple local and global optima correspond to distinct safety-critical situations \cite{ding2023survey,ding2021multimodal,wu2024lambda,klischat2019generating}. 

Most literature has primarily focused on identifying a single optimum of the mentioned optimization problem, corresponding to either the most safety-critical driving scenario or a predefined critical case \cite{wang2021advsim,klischat2019generating}.
However, focusing on the most safety-critical scenario or a predefined case cannot ensure safety across all circumstances and provide a comprehensive evaluation of AVs. 

Several studies \cite{jonkman_generation_2021,ding2021multimodal,wu2024lambda,wang2022comprehensive,tian_mosat_2022} address multimodal optimization problems to identify as many challenging and safety-critical scenarios as possible. \cite{ding2021multimodal} solves the challenge of estimating a multimodal distribution of safety-critical scenarios by using a flow-based generative model optimized within an on-policy optimization framework. Furthermore, \cite{wu2024lambda} addresses a multi modal optimization problem using an adaptive sampling method to identify and cover critical failure scenarios within a logical scenario space. Moreover, \cite{wang2022comprehensive} suggests an interaction-aware safety evaluation methodology that generates customized, challenging scenarios for AVs using an optimization-driven adaptive scheme. The studies
\cite{jonkman_generation_2021,tian_mosat_2022} focus on optimizing for diverse safety-critical scenarios by incorporating a novelty objective,  which quantifies how different a scenario is from previously encountered scenarios based on the average distance, in a multi-objective optimization framework.



Identifying all local and global optima that maximize the criticality metric for a comprehensive assessment of safety-critical scenarios is computationally demanding and costly. Therefore, the aforementioned studies  \cite{ding2021multimodal,wu2024lambda,jonkman_generation_2021,wang2022comprehensive,tian_mosat_2022}, suggest approaches to cover different modes of risky scenarios and achieve comprehensive coverage and a diverse set of safety-critical cases. However, these approaches generally rely on an archive of predefined test scenarios, constraining the exploration space to predetermined scenarios. Moreover, diversity is generally assessed using scenario parameters or ad hoc features, rather than customized challenges in control and system design, such as  features of the System Under Test (SUT) in the most safety-critical driving scenes. In addition, an explicit definition of diversity within the domain of driving scenarios is typically not considered \cite{braun2023review}.
	
In this paper, we employ angle-based features, defined by angle variables that characterize different geometric configurations at the most safety-critical driving scenes. We further define a diversity metric based on these angle-based features rather than system parameters. To achieve a comprehensive assessment of safety-critical scenarios, we identify a set of optima that are diverse in these features. The concept of diversity is essential for differentiating driving scenarios based on a dissimilarity metric \cite{ricotta2021towards}.
The dissimilarity metric quantifies differences between two scenarios, whereas the diversity metric
represents how comprehensively the feature space is covered. 
Therefore, we incorporate a dissimilarity metric into its search optimization algorithm to cover different operating conditions characterized by distinct angle-based features. 
	
	In general, dissimilarity is a metric evaluating two driving scenarios based on their differences. These differences can be in scenario parameter values \cite{tuncali2019rapidly,zhu2021hazardous}, in entire trajectory information using trajectory similarity metrics \cite{ries2021trajectory,nguyen2021salvo}, or scenario features extracted from trajectory information \cite{kerber2020clustering,kruber2018unsupervised,wheeler2019critical}. A parameter-based dissimilarity metric is defined to evaluate differences in scenario parameters and VOIs. In contrast, a feature-based dissimilarity metric distinguishes between two scenarios based on a set of feature variables derived from VOIs, including, but not limited to, vehicle position, orientation, speed, and acceleration. Focusing on angle-based features rather than on scenario parameters allows comparing two scenarios based on differences at the instant of maximum criticality \cite{mahadikar2024formulating}. Therefore, such an approach compares only the most relevant parts of scenarios from a safety verification standpoint, effectively eliminating redundancies. 
	
As mentioned previously, incorporating diversity and criticality into the optimization framework results in a multi-model optimization problem for scenario-based testing. To solve such optimization problems, evolutionary algorithms are advantageous, as they avoid analytic gradients and reduce the cost of function evaluation \cite{zingg2008comparative}. Moreover, niching techniques can be incorporated into evolutionary methods to maintain population diversity and discover multiple optimal solutions within a single population \cite{mahfoud1995niching, shir2005niching, li2016seeking}. Therefore, we employ a species-based niching method, the Species Conserving Genetic Algorithm (SCGA), introduced in \cite{li2002species}. This method preserves diversity in the decision space by identifying and preserving species, groups of individuals in a population that 
	have similar characteristics based on a Euclidean distance dissimilarity metric, ensuring that the population explores multiple optima rather than converging prematurely to a single solution. 
	
	To extend diversity  in angle-based features for safety-critical scenarios, we adapt SCGA by incorporating a feature-based clustering strategy. Specifically, we integrate the feature-based dissimilarity metric proposed in \cite{mahadikar2024formulating} into the SCGA search strategy, resulting in a feature-speciation evolutionary algorithm. 
	For comparison, we also implement a parameter-speciation approach by applying the traditional SCGA formulation in our scenario-based testing setting. The feature-speciation approach explicitly identifies safety-critical scenarios with diversity in angle-based features, while the parameter-speciation approach identifies safety-critical scenarios based on diversity in scenario parameters. Therefore, diversity in angle-based features is not guaranteed by the parameter-speciation approach. As a result, parameter-speciation may yield scenarios that appear diverse in the parameter space but remain similar in angle-based features at the most safety-critical moment.
	
	To evaluate the effectiveness of both methods, we represent an explicit definition of scenario diversity, called feature-based diversity measure. This measure is based on the feature-based dissimilarity metric that quantifies the diversity of the generated scenarios with respect to features. 
	The performance of the algorithms in generating diverse sets of safety-critical scenarios is  demonstrated through a simulation-based use case. Accordingly, the main contributions of the paper are as follows:
 \nj{	
	\begin{itemize}
		\item Using the scenario-based testing definitions in \cite{kevin}, we formulate scenario generation as a multimodal optimization problem based on the behavioral modeling approach.
		\item We propose a feature-speciation evolutionary algorithm that systematically identifies a set of diverse safety-critical scenarios within the feature space.
		\item We evaluate the feature-speciation method in comparison with the parameter-speciation approach using a feature-based diversity measure. 
	\end{itemize}  
}	
	The remainder of the paper is organized as follows. Section \ref{sec:pre} provides essential background on the definitions of scenario, criticality metrics, and dissimilarity metrics. The problem statement is introduced in Section~\ref{sec:prob}, followed by the behavioral modeling formulation in Section~\ref{sec:Model},  the formulation of the scenario-based optimization problem in Section~\ref{sec:OPT}, and feature-based and parameter-based dissimilarity metric descriptions in Section~\ref{sec:diss}. The optimization search algorithm is presented in Section~\ref{sec:alg}, followed by the diversity-driven optimization objective in Section~\ref{sec:divprob}, the feature-speciation evolutionary algorithm in Section~\ref{sec:alg-f}, the parameter-speciation evolutionary algorithm in Section~\ref{sec:alg-p}, and the feature-based diversity measure in Section~\ref{sec:divf}. The performance of the proposed method is demonstrated in Section \ref{sec:exa} using a use case. Section \ref{se:con} concludes the paper.

	\textbf{Notation:}
	$\mathbb{T} \subseteq \mathbb{R}_{\geq 0}$ denotes an open subset of the non-negative real numbers. 
	Moreover, $\mathbb{R}_{> 0} \subset \mathbb{R}$ denote  the set of positive real numbers, respectively.

	\section{Preliminaries}
	\label{sec:pre}
	In this section, we introduce the concepts of scenario definition, criticality metrics, and dissimilarity metrics.

	\subsection{Scenario Description}
	\label{sec:scenario}
	
	In this section, we introduce the mathematical formulation of scenarios and related terminology in scenario-based testing using the behavioral modeling approach presented in \cite{kevin} as follows:
	\begin{itemize}
		\item \emph{System of Interest $\Sigma$}: The system of interest refers to a real-world physical system whose behavior is to be studied or modeled. 
		\item  \emph{Behavior and Behavioral Relations}: A behavior $\mathfrak{B}$ is a subset of trajectories of Variables of Interest (VOIs) over time that satisfy specific conditions, known as behavioral relations. These behavioral relations are a set of mathematical equations that define the  behavior of dynamics of the system of interest.
		\item  \emph{Objects} $O_i$: An object $O_i$, alternatively called an element $O_i$, is a physical entity of interest from a modeling perspective, such as a vehicle, a traffic light or a road marking.
		\item  \emph{Variables}: Variables are quantities that can vary over time and are categorized into two groups:
		\begin{enumerate}
			\item VOIs $\zeta(t)$ are variables of modeling interest, such as, e.g., the position $(x_i(t),y_i(t))$ of $O_i$ in a two-dimensional space.
			\item  Latent variables $\alpha(t)$ represent auxiliary variables that define behavioral relationships beyond VOIs, such as velocities $({{v}_x}_i(t),{{v}_y}_i(t))$ and accelerations $({{a}_x}_i(t),{{a}_y}_i(t))$ of $O_i$. 
		\end{enumerate}
		\item  \emph{System parameters} $\eta$: System parameters are all constants appearing in the behavioral relations, including initial conditions for each VOI, such as, e.g., $(x_i(0),y_i(0))$, and initial conditions  for latent variables, such as, e.g., $({{v}_x}_i(0),{{v}_y}_i(0))$ and $({{a}_x}_i(0),{{a}_y}_i(0))$.
		\item \emph{Concrete Scenario and Scene}:
		A concrete scenario is a specific trajectory of VOIs within the behavior of the system, defined over a specified time interval, which is a subset of $\mathbb{T}$. A scene represents a specific time instant of a concrete scenario.
	\end{itemize} 
	\subsection{Criticality Metrics}
	\label{sec:cm}
	Criticality metrics, which measure the behavioral safety of a vehicle, have been introduced in \cite{westhofen2023criticality} as mappings from the set of all traffic scenes to a subset of the real numbers. These metrics can be categorized into scene-level metrics, which are computed using information at a specific point in time, and scenario-level metrics, calculated over a given time series.

	\subsection{Dissimilarity Metrics}
	
	A dissimilarity metric quantifies the difference between two driving scenarios based on specific characteristics. In general, it can be defined as a function $ \mathfrak{d}: \mathfrak{B} \times \mathfrak{B} \to \mathbb{R}_{> 0} $, which compares two scenarios within the same behavior set $ \mathcal{B}$. Depending on the application, this metric can be based on the VOIs or on feature variables that are functions of the VOIs \cite{mahadikar2024formulating}. 
	\section{Problem Statement}
	\label{sec:prob}
	In this section, we first formalize the definition of scenarios and related concepts introduced in Section \ref{sec:pre} using the behavioral modeling approach. Based on the behavioral modeling concept, we introduce the feature-based dissimilarity metric used for clustering diverse scenarios. Finally, we formulate an optimization problem for generating safety-critical scenarios. 
	
	\subsection{Behavioral Modeling Formulation}
	\label{sec:Model}
	
	Consider a traffic system of interest, denoted by $\Sigma$, composed of elements $O_e$ and $O_1$,
	representing the ego actor and the adversarial actor $i$, respectively. We define the behavioral set of $\Sigma$ as the set of all admissible trajectories of the VOIs over the time interval $\mathbb{T}$. A trajectory is considered admissible if it satisfies the behavioral relations. These behavioral relations are defined by the safety, physical, and operational limits represented by the constraint function $C_1$, and by the vehicle dynamics represented by the constraint function $C_2$, typically expressed through differential operators.
		The behavioral set can therefore be defined as follows:
		\begin{equation}
			\label{eq:beh}
			\mathfrak{B} = \Big\{
			\zeta : \mathbb{T} \to \mathbb{R}^n \;\Big|\;
			\begin{aligned}
				& C_1(\zeta(t), \eta, \alpha(t),D\alpha(t)) \preceq \mathbf{0}, \\
				& C_2(\zeta(t), \eta, \alpha(t),D\alpha(t)) = \mathbf{0}
			\end{aligned}
			\Big\}.
		\end{equation}
		where $\zeta(t)\in \mathbb{R}^n$ and $\eta \in \mathbb{R}^r$, with $n,r\in\mathbb{N}$ are positive integers, representing a vector of $n$ VOIs and a vector of $r$ system parameters of $\Sigma$, respectively. The latent variables are given by $\alpha: \mathbb{T} \to \mathbb{R}^m$, where $m \in \mathbb{N}$ denotes the number of latent variables. Moreover,  $D:=\frac{d}{dt}(\cdot)$ denotes the differentiation operator. The functions $C_1: \mathbb{R}^n \times \mathbb{R}^r \times \mathbb{R}^{m} \times \mathbb{R}^m \to \mathbb{R}^{q_1}$ and $C_2: \mathbb{R}^n \times \mathbb{R}^{r} \times \mathbb{R}^m \times \mathbb{R}^m \to \mathbb{R}^{q_2}$, where $q_1,q_2 \in\mathbb{N}$, define the behavioral relations of $\Sigma$.

	Therefore, a concrete scenario $\zeta_s(t)$ over a specific time interval $\mathbb{T}_{s}$ is described as the following well-defined function:
	
	\begin{equation}
		\label{eq:scenario}
		{\zeta_{s}}: \mathbb{T}_{s} \to \mathbb{R}^n, \quad \zeta_s(t) \in \mathfrak{B} \quad \forall t \in \mathbb{T}_{s},
	\end{equation}
	where $\mathbb{T}_{s} \subseteq \mathbb{T}$. Moreover,
	a scene is a vector containing the values of all VOIs at a specific time instant within the behavior of the system. According to the definition of the concrete scenario given in \eqref{eq:scenario}, a scene $\zeta_{s}^{\star}$ at a given time $t^\star$ is defined as follows:  
	\begin{equation}
		\label{eq:scene}
		\zeta_{s}^{\star}:=\zeta_{s}(t^\star),
	\end{equation} 
	where $t^\star \in \mathbb{T}_{s}$.
	A scenario-level criticality metric $\mathfrak{C}(\zeta_s(t))$ is defined as follows:
		\vspace*{-2mm} 
	\begin{equation}
		\mathfrak{C}: \mathfrak{B} \to \mathbb{R}_{\geq 0} ,
	\end{equation}
	
	A scene-level metric $\mathfrak{C}(\zeta_{s}^{\star})$ at a given time $t^{\star}$ can be defined as follows:
	
	\begin{equation}
		\mathfrak{C}: \mathbb{R}^{n} \to \mathbb{R}_{\geq 0} , \quad \mathfrak{C}(\zeta_{s}(t^\star))=\mathfrak{C}(\zeta_{s}^{\star}).
	\end{equation}
	
	In this paper, we use a distance-based criticality metric, which can be defined as follows:

		\begin{equation}
			\label{eq:Critical}
			\mathfrak{C}(\zeta_{s}) = 
			\begin{cases}
				+\frac{1}{2} v_{\text{col}},  \qquad  \qquad  \quad \text{ if } d_{\text{min}} = 0 \\
				-\min\limits_{t_{k} \in \mathbb{T}_s}d_{\text{min}}(t_k), \qquad \text{otherwise}
			\end{cases}
		\end{equation}
		with
		\begin{equation}
			d_{\text{min}}(t_{k})=\min \left\lVert q_1(\zeta_s(t_k)) - q_e(\zeta_s(t_k)) \right\rVert,
		\end{equation}
	where ${\zeta_{s}}(t_k)$ is a scenario at a specific time $t_k$ over the time interval $\mathbb{T}_s \subseteq \mathbb{T}$, and $q_1$ and $q_e$ are the positions of the adversarial actor
	and the ego actor, respectively. This criticality metric represents the minimum distance between the ego actor and the closest adversarial actor  during the scenario. The most critical time $t_c$ is defined 
	as the time at which this minimum distance occurs:
		\begin{equation}
			\label{eq:tc}
			t_c := \arg\min\limits_{t_{k} \in \mathbb{T}_s}d_{\text{min}}(t_k).
		\end{equation}
	In the case of a collision (i.e., $\min {\left\lVert {q}_{1}-{q}_{e} \right\rVert}=0$) at $t_c$, the criticality metric in \eqref{eq:Critical} is extended to incorporate the relative collision velocity $v_{\text{col}}$, which is used to score the scenario's criticality, as follows:   
	\begin{equation}
		\label{eq:psds}
		v_{\tt{col}}=\left\lVert{{p}_{1}({\zeta_{s}}(t_c))-{p}_{e}({\zeta_{s}}(t_c))}\right\rVert,
	\end{equation}  
	where $p_1$ and $p_e$ represent the velocities of the adversarial and the ego actors, respectively.
	\nj{
	\begin{remark}
		The criticality metric in \eqref{eq:Critical} is formulated based on the minimum distance between the ego and adversarial vehicles, modeled as point masses. This formulation facilitates
		a basis for studying diversity in safety-critical scenario generation through an optimization framework. Moreover, it can be extended by using more precise definitions of the minimum distance that account for the geometric dimensions and shapes of the vehicles. Alternatively, it can be generalized to incorporate probabilistic or risk-weighted criticality metrics.
	\end{remark}
}

	\subsection{Optimization Problem Formulation}
	\label{sec:OPT}
	In this section, we formulate an optimization problem for generating safety-critical scenarios by maximizing the criticality metric \eqref{eq:Critical} as the cost function. We consider the vector of system parameters $\eta \in \mathcal{E} \subseteq \mathbb{R}^r$ as the optimization variables, where $ \mathcal{E}$ denotes a  set of admissible vectors of system parameters. We assume that for each $\eta$ the corresponding scenario $\zeta$ is uniquely determined from the behavioral model \eqref{eq:beh}.
	
For notational convenience in describing the parametrization with respect to $\eta$, we define the parameterized behavioral set as 
	\begin{equation}
		\label{eq:beh2}
		\mathfrak{B}^\eta = \Big\{
		\zeta^\eta : \mathbb{T} \to \mathbb{R}^n \;\Big|\;
		\begin{aligned}
			& C_1(\zeta^\eta(t), \eta, \alpha(t),D\alpha(t)) \preceq \mathbf{0}, \\
			& C_2(\zeta^\eta(t), \eta, \alpha(t),D\alpha(t)) = \mathbf{0}
		\end{aligned}
		\Big\}.
	\end{equation}
	where $\zeta^\eta$ is the admissible scenario $\zeta^\eta(t)$ corresponding to a given vector of system parameters $\eta$.

	\begin{remark}
		\label{re:c-hat}
		Note that the behavioral set $\mathfrak{B}^\eta$ in \eqref{eq:beh2} explicitly characterizes the admissible scenario $\zeta^\eta(t)$ corresponding to a given vector of system parameters $\eta$, where the relation between $\zeta^\eta(t)$ and $\eta$ is governed by the constraints ${C}_1(\cdot) \preceq \mathbf{0}$ and ${C}_2(\cdot) = \mathbf{0}$. In detail, for a fixed $\eta$, the function $\zeta^\eta(t)$ is chosen such that the system dynamics and conditions denoted by ${C}_1$ and ${C}_2$ are satisfied, making $\zeta^\eta(t)$ a feasible scenario corresponding to the given system parameters. Moreover, we assume that the scenario $\zeta^{\eta}(t)$ is unique for each $\eta$.
	\end{remark}

	According to \eqref{eq:scenario}, the concrete scenario function parametrized by $\eta$ is denoted as follows
	\begin{equation}
		\label{eq:scenario2}
		{\zeta^\eta_{s}}: \mathbb{T}_s \to \mathbb{R}^n, \quad \zeta^\eta_s(t) \in 	\mathfrak{B}^\eta.
	\end{equation}
	
	The criticality metric \eqref{eq:Critical} is obtained as a function of $\eta$ by evaluating it at the parameterized scenario $\zeta^\eta_{s}(t)$ in \eqref{eq:scenario2}. Hence, we formulate the following optimization problem:
	\begin{equation}
		\label{eq:OPT}
		\hat{\eta} = \arg\max_{\eta} 
		\mathfrak{C}(\zeta^\eta_{s}), 
		\quad \text{s.t.} \quad \zeta^\eta_s \in 	\mathfrak{B}^\eta,
	\end{equation}
	where each $\hat{\eta} \in \hat{\mathcal{E}} \subseteq \mathbb{R}^r$ is a vector of $r$ system parameters that maximizes the criticality metric. Moreover $ \hat{\mathcal{E}}$ denotes a finite set of critical system parameter vectors $\hat{\eta}$ obtained through this multimodal optimization, as precisely introduced in Section \ref{sec:divprob}. The cost function $\mathfrak{C}$ is defined using the criticality metric from \eqref{eq:Critical}, evaluated at $\zeta^\eta_{s}(t)$.  
	Note that the mapping from $\eta$ to $\zeta^\eta_{s}(t)$ is implicitly governed by the constraints in \eqref{eq:beh2}.  Therefore, $\hat{\eta}$ yields the critical scenario $\zeta^{\hat{\eta}}_{s}(t)$ that characterize (near-)collision circumstances.
	
	\begin{remark}
		In this study, we focus on critical scenarios involving an ego vehicle and adversarial actors. Therefore, critical scenarios between other road elements and actors are not considered. However, the framework can be extended to more general scenarios by incorporating additional road users and interaction types into the behavioral modeling and  optimization formulation. 
	\end{remark}
	
	\subsection{Parameter-based and feature-based dissimilarity metrics}
	\label{sec:diss}
	In this section, we introduce two dissimilarity metrics--- parameter-based and feature-based--- for comparing two distinct scenarios within the same behavior set, $\mathfrak{B}^\eta_{\bar\alpha}$. \nj{The parameter-based metric compares two driving scenarios based on VOIs across their full trajectories, such as velocity.  Conversely, the feature-based metric compares two driving scenarios based on specific features extracted at the most critical scene, such as the heading angle difference. }
	These metrics provide the foundation of the optimization algorithms developed in the subsequent sections, which aim to identify diverse sets of critical scenarios.
	
	\nj{Let $\zeta^{\eta^{(\nu)}}_{s}(t)$ and $\zeta^{\eta^{(\ell)}}_{s}(t)$ denote two scenarios within the same behavioral set $\mathfrak{B}^\eta_{\bar\alpha}$, corresponding to two distinct and admissible system parameter vectors $\eta^{(\nu)},\eta^{(\ell)} \in \mathcal{E} $. Suppose that the features $f^{(\nu)},f^{(\ell)} : \mathbb{R}^n \to \mathbb{R}^q  $ are defined as scenario-dependent functions, where their input represent the most critical scene in the respective scenarios, $\zeta^{\eta^{(\nu)}}_{s}(t)$ and $\zeta^{\eta^{(\ell)}}_{s}(t)$.  
	
	We define the parameter-based dissimilarity metric $\mathfrak{d}_p$ as the normalized Euclidean distance between the parameter vectors:} 
	 \begin{equation}
	 	\label{eq:diss2}
	 	\mathfrak{d}_{p}(\zeta^{\eta^{(\nu)}}_{s}, \zeta^{\eta^{(\ell)}}_{s})
	 	\footnote{Because each scenario is uniquely determined by its system parameter vector, we abbreviate them as $\zeta^{\eta^{(\nu)}}_{s}$ and $\zeta^{\eta^{(\ell)}}_{s}$.}
 	:= \sqrt{\frac{1}{r}\sum_{m=1}^r (\eta^{(\ell)}_{m} - \eta^{(\nu)}_{m})^2},
	 	\end{equation}
 	   	where ${\eta^{(\nu)}} = [\eta^{(\nu)}_{1}, \dots, \eta^{(\nu)}_{r}]^\top$ and ${\eta^{(\ell)}} = [\eta^{(\ell)}_{1}, \dots, \eta^{(\ell)}_{r}]^\top$
 	    represent $r$-dimensional system parameter vectors.
	 \nj{As defined in \eqref{eq:diss2}, a greater difference between the system parameters yields a higher parameter-based dissimilarity $\mathfrak{d}_p$. }  
	 
	 \nj{The feature-based dissimilarity metric can be defined as the weighted Euclidian distance between selected features at the most critical scene in each scenario:
	 \begin{equation}\label{eq:diss}
	 		\mathfrak{d}_{f}(\zeta^{\eta^{(\nu)}}_{s}(t_c^\nu), \zeta^{\eta^{(l)}}_{s}(t_c^l)):=
	 		\sqrt{\sum_{n=1}^q w_n (f^{(\ell)}_{n} - f^{(\nu)}_{n})^2},
	 \end{equation}
	where ${f^{(\nu)}} = [f^{(\nu)}_{1}, \dots, f^{(\nu)}_{q}]^\top$ and ${f^{(\ell)}} = [f^{(\ell)}_{1}, \dots, f^{(\ell)}_{q}]^\top$ denote $q$-dimensional feature vectors 
	 evaluated at the most critical moments $t_c^\nu, t_c^\ell \in \mathbb{T}_s$, respectively. The positive constants $w_n \in \mathbb{R}^{+}$ represent the weights assigned to each features.}

\begin{figure}[t]
		\centering
		\includegraphics[trim= 0.1in 0 0.1in 0, clip, width = 0.4\linewidth]{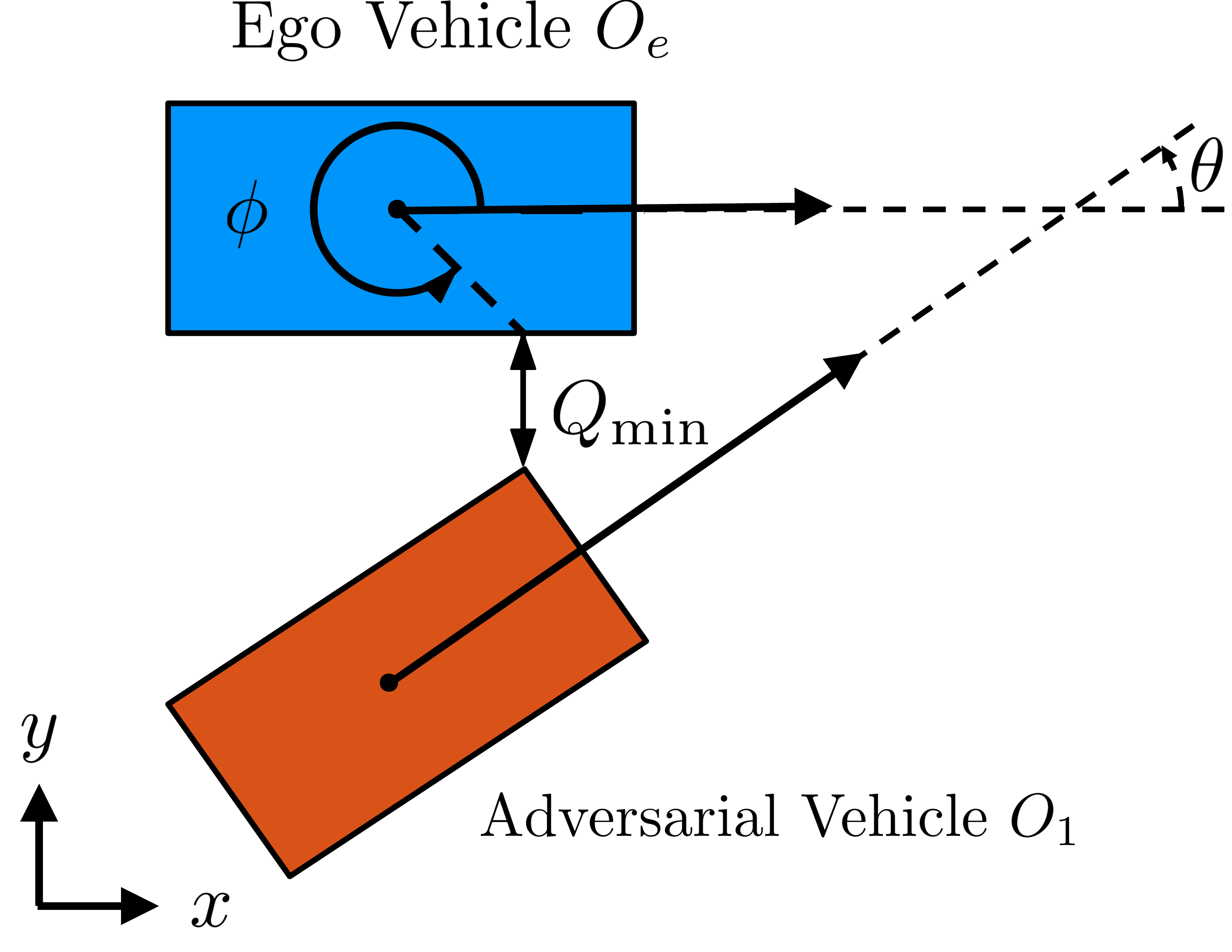}
		\caption{The angle-based features: The heading angle difference $(\theta)$, defined as the angle between the heading directions of the ego vehicle $O_{e}$ (blue) and the adversarial vehicle $O_{1}$ (red); and the angular position in the ego frame $(\phi)$, defined as the angle from the the x-axis of the ego frame to the vector from $O_{e}$ to the closest point on $Q_{1}$, denoted as $Q_{\min}$. }
		\label{fig:angular feature}
	\end{figure}

	\nj{As proposed in \cite{mahadikar2024formulating}}, we employ two angle-based features $\theta: \mathbb{R}^n \to [-\pi,\pi)$ and $\phi: \mathbb{R}^n \to [-\pi,\pi)$  that characterize the geometric configuration of (near-) collision between the ego and adversarial vehicles, as illustrated in Fig.~\ref{fig:angular feature}. 
	
 These features are detailed as follows:
	\begin{itemize}
		\item Heading Angle Difference $\theta(\cdot)$: This feature represents the difference in heading angles between the ego vehicle and the adversarial actor at the most critical moment\footnote{The most critical moment $t_{\text{c}}$ represents the time at which the minimum distance in \eqref{eq:Critical} occurs between the adversarial and ego actors within the scenario, as mentioned in \eqref{eq:tc}.
		}, $t_{c} \in \mathbb{T}$, in a given scenario $\zeta^{\eta}_{s}(t)$. It helps distinguish between critical scenarios involving adversarial actors moving in different directions with respect to the ego vehicle-- for example, a head-on collision versus a rear-end collision.
		\item Angular Position in Ego Frame $\phi(\cdot)$:
		This feature represents the angular position of the adversarial actor relative to the ego vehicle frame of reference at $t_{c}$, in a given scenario $\zeta^{\eta}_{s}(t)$.
		It helps distinguish between different collision scenarios by indicating where the collision occurs on the ego vehicle-- for example on the front, rear or side.
	\end{itemize}	
	The absolute angular differences between two scenarios are computed as follows\footnote{To simplify notation, we omit the arguments from $\theta^{(\nu)},\theta^{(\ell)}$ (i.e., $\theta^{(\nu)}(\zeta^{\eta^{(\nu)}}_{s}(t_c^\nu))$, $\theta^{(\ell)}(\zeta^{\eta^{(\ell)}}_{s}(t_c^\ell))$) and $\phi^{(\nu)},\phi^{(\ell)}$ (i.e., $\phi^{(\nu)}(\zeta^{\eta^{(\nu)}}_{s}(t_c^\nu))$, $\phi^{(\ell)}(\zeta^{\eta^{(\ell)}}_{s}(t_c^\ell))$) in the formulations. }:
	\begin{eqnarray}
		&\Delta{\theta} := \min{(|\theta^{(\nu)} - \theta^{(\ell)}|,|2\pi - (\theta^{(\nu)} - \theta^{(\ell)})| )},	\label{eq:feature1}\\&
		\Delta{\phi} := \min{(|\phi^{(\nu)} - \phi^{(\ell)}|,|2\pi - (\phi^{(\nu)} - \phi^{(\ell)})|)},
		\label{eq:feature2}
	\end{eqnarray}
	where $\theta^{(\nu)}, \phi^{(\nu)} \in \mathbb{R}$ are the features of scenario $\zeta^{\eta^{(\nu)}}_{s}(t) $ and $\theta^{(\ell)} , \phi^{(\ell)} \in \mathbb{R}$ represent the features of the scenario $\zeta^{\eta^{(\ell)}}_{s}(t)$. 
	
	\nj{According to \eqref{eq:diss},} the feature-based dissimilarity  metric based on the angle-based features is defined as follows:
	\begin{equation}\label{eq:diss3}
		\mathfrak{d}_{f}(\zeta^{\eta^{(\nu)}}_{s}(t_c^\nu), \zeta^{\eta^{(l)}}_{s}(t_c^l))= \sqrt{w_1\Delta{\theta}^2+w_2\Delta{\phi}^2},
	\end{equation}

	The weights  $w_1, w_2 \in \mathbb{R}^{+}$ can be adjusted based on the specific use case. For instance, in highway scenarios, differences in heading angle are not very significant. Therefore, smaller values for $w_1$ can be appropriate. 
	
		\nj{\begin{remark}
		The proposed parameter-based and feature-based dissimilarity metrics in \eqref{eq:diss2} and \eqref{eq:diss} can be extended or replaced by other formulations, while preserving the dissimilarity metric definition.  Moreover, the features selected to compute the metric in \eqref{eq:diss} or the vector of system parameter $\eta$ can be customized to the operational requirements and specific challenges.  In this work, the two angle-based features, $\theta$ and $\phi$, are used to characterize the geometric configuration of a potential collision, as suggested in \cite{mahadikar2024formulating}.	
	\end{remark}}

	In the next sections, we focus on developing an algorithm to solve the  optimization problem \eqref{eq:OPT}, and on how the feature-based dissimilarity metric \eqref{eq:diss} can be integrated into the search algorithm to identify a diverse set of safety-critical scenarios with respect to the features.
	
	\section{Optimization Search Algorithm }
	\label{sec:alg}
	To solve the optimization problem \eqref{eq:OPT} and generate a diverse set of safety-critical scenarios, the following challenges are considered:
	\begin{itemize}
		\item The cost function $\mathfrak{C}(\zeta^\eta_{s}(t))$ in \eqref{eq:OPT} may not be convex nor possess an analytic gradient.
		\item Finding multiple (local/global) maxima rather than a single (or global) maximum in \eqref{eq:OPT} enables generating all possible safety-critical scenarios within the ODD.	
		\item Generating diverse safety-critical scenarios with respect to a specific measure, i.e., identifying a diverse set of optima within the solution space, is crucial for enhancing ODD coverage.
		\item 
		To discover scenarios that differ in the most safety-relevant scenes, the optimization algorithm needs to preserve diversity among multiple optima in the feature space.
	\end{itemize}
	
	To address the mentioned challenges, we adopt the  Species Conserving Genetic Algorithm (SCGA) approach to the optimization problem \eqref{eq:OPT}. This method introduces an evolutionary algorithm that identifies multiple optimal solutions. It also preserves  diversity among the population by clustering the population into several sub-populations, called species, and maintaining these species in parallel throughout the evolutionary process.
	Species are formed based on the similarity of individuals in the population, clustered around a dominant individual known as the species seed. The species seeds identified in the current generation are then preserved by moving  them over to the next generation.  These species seeds represent candidate solutions converging toward multiple optima \cite{li2002species}. 
	
	In traditional SCGA, diversity is preserved in the parameter space using a distance-based dissimilarity metric. In our scenario-based testing setting, we refer to this as the parameter-speciation approach, where species are formed based on the parameter-based dissimilarity metric \eqref{eq:diss2}. To preserve diversity in the feature space instead, we develop a feature-speciation approach, which relies on the feature-based dissimilarity metric \eqref{eq:diss}. 
	
	In the following sections, we first translate the scenario-based testing challenges into an optimization objective. We then introduce both the feature- and parameter-speciation approaches and compare their performance using a feature-based diversity measure.
	\subsection{Diversity-Driven Optimization Objective}
	\label{sec:divprob}
	
	The optimization problem in \eqref{eq:OPT} may admit multiple solutions, 
	including both global and local optima of the cost function $\mathfrak{C}(\zeta^\eta_{s}(t))$. We denote the optimal solutions by the critical system parameter $\hat{\eta}$. The set of all possible optimal values is represented by $\hat{\mathcal{E}} \subseteq \mathbb{R}^r$ and is defined as follows.
	
	
	\begin{definition}
		The set of optimal solutions of \eqref{eq:OPT} is defined as
		\begin{equation}
			\label{eq:local_optima_set}
			\hat{\mathcal{E}} := \left\{ \hat{\eta} \in \mathbb{R}^r \ \middle| \
			\begin{aligned}
				&\exists\, \epsilon > 0 \text{ such that } \forall \eta \in \mathbb{R}^r, \\
				&\|\eta - \hat{\eta}\| \leq \epsilon \ \Rightarrow \
				\mathfrak{C}(\zeta_s^{\hat{\eta}}) \geq \mathfrak{C}(\zeta_s^\eta)
			\end{aligned}
			\right\},
		\end{equation}
		where $\zeta_s^\eta(t) \in \mathfrak{B}^\eta$ and 
		$\zeta_s^{\hat{\eta}}(t) \in \mathfrak{B}^{\hat{\eta}} := \mathfrak{B}^\eta|_{\eta = \hat{\eta}}$, 
		and $\epsilon > 0$ defines a local neighborhood around $\hat{\eta}$.
	\end{definition}

	However, the optimal values in $\hat{\mathcal{E}}$ may correspond to critical scenarios with identical angle-based features $\theta$ and $\phi$. Therefore, the optimization objective is to identify a subset $\hat{\mathcal{E}}_{\text{div}} \subseteq \hat{\mathcal{E}}$ consisting of parameter values $\hat{\eta}$ that yield safety-critical scenarios that are diverse in terms of angle-based features within the ODD.
	
	To formalize the notion of diversity, we introduce a feature-based diversity measure, inspired by Rao’s Quadratic Diversity \cite{rao1982diversity}.  
	
	\begin{definition}[Feature-based Quadratic Diversity]
		\label{def:Q}  
		Suppose the set of $M \in \mathbb{N}$ optimal parameter vectors in \eqref{eq:local_optima_set} is given by   
		\begin{equation}  
			\label{eq:hatE}  
			\hat{\mathcal{E}} = \{\hat{\eta}^{(1)},\dots, \hat{\eta}^{(M)}\}.  
		\end{equation}  		
		The feature-based quadratic diversity measure is defined as    
		\begin{equation}  
			\label{eq:div}  
			\mathcal{Q} \;=\; \frac{1}{M(M-1)}\sum_{i=1}^{M}\sum_{j=1}^{M} \mathfrak{d}_{f}\!\left( \zeta^{\hat{\eta}^{(i)}}_{s}(t_c^i), \, \zeta^{\hat{\eta}^{(j)}}_{s}(t_c^j)\right),  
		\end{equation}  
		where $\zeta^{\hat{\eta}^{(i)}}_{s}(t_c^i)$ and $\zeta^{\hat{\eta}^{(j)}}_{s}(t_c^j)$ are the critical scenarios corresponding to $\hat{\eta}^{(i)}$ and $\hat{\eta}^{(j)}$ at the most critical moments $t_c^i, t_c^j \in \mathbb{T}_s$. The dissimilarity metric $\mathfrak{d}_{f}$ is defined in \eqref{eq:diss} and satisfies $\mathfrak{d}_{f}=0$ whenever $i=j$.  
	\end{definition}
	
	This metric represents the mean pairwise feature-dissimilarity among  the set of critical scenarios. A larger value of $\mathcal{Q}$ indicates that the scenarios are more widely spread and therefore more diverse in the feature space.  
	\subsection{A feature-speciation evolutionary algorithm}
	\label{sec:alg-f}
	To identify the subset $\hat{\mathcal{E}}_{\text{div}}$, we propose a modified version of the SCGA, as presented in Algorithm~\ref{alg:SCGA}.
	\renewcommand{\algorithmicrequire}{\textbf{Input:}}
	\renewcommand{\algorithmicensure}{\textbf{Output:}}
	\begin{algorithm}
		\caption{Feature-based SCGA  }\label{alg:SCGA}
		\begin{algorithmic}[1]
			\Require{Generation index $k$, indicating the $k$-th generation $G_{k}$; Population size $N$; Individual $\eta_{k}^{(i)}$ representing the $i$-th system parameter candidate in generation $k$; Population at $G_{k}$: $P_{k} := \{ \eta_k^{(i)}\}_{i=1}^N$; Maximum number of generations $\text{MaxG}$; Dissimilarity metric $\mathfrak{d}_f$ in \eqref{eq:diss};  Dissimilarity threshold $\sigma_s$.}
			\Ensure{A set of diverse safety-critical scenarios ${\zeta^{\hat{\eta}}_{s}}(t)$.}
			\State {Initialize the generation index $k$, that is, $k \gets 0$.}
			\State {Initialize population $P_{k}$ for generation $G_{k}$.}
			
			
			
			\State {Evaluate the criticality values $\mathfrak{C}(\zeta^{\eta^{(i)}_k}_s)$, as defined in~\eqref{eq:Critical}, for the scenario trajectories $\zeta^{\eta^{(i)}_k}_s(t)$ of each individual $\eta^{(i)}_k \in P_{k}$ at $G_{k}$, using Remark~\ref{re:c-hat}.}
			\While{k $\le$ MaxG}
			\State Identify the set of species seeds $X_{{S}_k}$ in generation $G_{k}$ by clustering the population $P_{k}$ into multiple species using the feature-based dissimilarity metric $\mathfrak{d}_f$, as described in \textit{Algorithm~\ref{alg:identification}}.
			
			\State Select individuals $\eta_k^{(i)} \in P_{k}$ for reproduction to form the mating pool for generation $G_{k+1}$.
			
			\State Apply crossover and mutation within each species to generate offspring $\eta^{(i)}_{k+1}$ and form the new population $P_{k+1}$ for generation $G_{k+1}$.
			
			\State Evaluate criticality values $\mathfrak{C}(\zeta^{\eta_{k+1}^{(i)}}_s)$ of trajectories $\zeta^{\eta_{k+1}^{(i)}}_s(t)$ for all individuals $\eta_{k+1}^{(i)} \in P_{k+1}$, using Remark \ref{re:c-hat}.		
			\State Conserve species from $X_{{S}_k}$ in generation $G_{k+1}$ using the feature-based species conservation \textit {Algorithm~\ref{alg:feature_conservation}}.
			\State Update the generation index $k$, that is, $k \gets k + 1$.  
			
			\EndWhile	
		\State Repeat the species seed identification Step~5 for the final generation $G_{k}$, where $k = \text{MaxG}$.
		\State Store the set of final species seeds representing the critical system parameter values as
		$\hat{\mathcal{E}}_{\text{div}} = \{ \hat{\eta} \mid \hat{\eta} \in X_{{S}_k},\; k = \text{MaxG}+1 \}$.			
		\State Compute the safety-critical scenarios $\zeta^{\eta}_s(t)$ corresponding to each $\eta = \hat{\eta} \in \hat{\mathcal{E}}_{\text{div}}$ using Remark~\ref{re:c-hat}.
		\\
		\Return The diverse safety-critical scenarios set corresponding to the critical system parameter values $\hat{\eta}$ is given by $\{ \zeta^{\hat{\eta}}_s(t) \mid \hat{\eta} \in \hat{\mathcal{E}}_{\text{div}} \}$.	
	\end{algorithmic}
\end{algorithm}

In Algorithm~\ref{alg:SCGA}, we define a generational loop indexed by $k$, which aims to let the candidate individuals converge toward $\hat{\eta}\in \hat{\mathcal{E}} $ in \eqref{eq:local_optima_set} as $k$ increases. In each generation $k$ (i.e., $G_{k}$), a population $P_{k}$ consisting of $N$ individuals is considered, where each individual $\eta_k^{(i)}$ (for $i = 1,2, \dots, N$) represents a candidate system parameter value at that generation. This system parameter parametrizes the corresponding safety-critical scenario, denoted by $\zeta^{\eta_k^{(i)}}_s(t)$. The associated critical scenario is obtained through the process described in Remark~\ref{re:c-hat}. To determine the candidate individuals $\eta_{k+1}^{(i)} \in P_{k+1}$ transferring into the next generation $G_{k+1}$, we need to cluster the current population $P_{k}$ into a set of species and identify the dominant individuals in each species, as described in Step 5 of Algorithm \ref{alg:SCGA}. The dominant individuals are denoted by the set of species seeds $X_{{S}_k}$, which are identified in each generation $G_{k}$ using the species seed determination procedure described in Algorithm \ref{alg:identification}.   
\begin{algorithm}
	\caption{Feature-Based Species Seed Identification}
	\label{alg:identification}
	\begin{algorithmic}[1]
		\Require{Individual $\eta_k^{(i)}$ representing the $i$-th system parameter candidate in generation $k$; Population at generation $G_{k}$: $P_{k} := \{ \eta_k^{(i)}\}_{i=1}^N$; Trajectories $\zeta^{\eta_k^{(i)}}_s(t)$ and corresponding criticality values $\mathfrak{C}(\zeta^{\eta_k^{(i)}}_s)$ for all individuals $\eta_k^{(i)} \in P_{k}$, computed in Step 4 of Algorithm~\ref{alg:SCGA}; Dissimilarity metric $\mathfrak{d}$ in \eqref{eq:diss}; Dissimilarity threshold $\bar\sigma_f$.}
		\Ensure{Set of species seeds  $X_{S_k}$.}
		
		\State Initialize the set of species seeds $X_{{S}_k}$ at generation $k$, that is, $X_{S_k} \gets \emptyset$.
		\State Mark all individuals $\eta_k^{(i)} \in P_{k}$ as unprocessed. 
		
		\While{there exist unprocessed individuals $\eta_k^{(i)} \in P_{k}$}
		\State {Select the unprocessed individual with the highest criticality value $\mathfrak{C}(\zeta^{{\eta}^{(i)}_k}_s)$
			at the critical moment $t_c$ as ${\eta_k^\star}^{(i)} \in P_{k}$.}
		\State Mark ${{\eta_k}^\star}^{(i)} $ as processed.
		\State $found \gets \textbf{false}$.
		
		\For{all $x_{k} \in X_{S_k}$}
		\State Evaluate the absolute values and differences between the features $\theta_{{{\eta_k}^\star}^{(i)}},\phi_{{{\eta_k}^\star}^{(i)}}$ and $\theta_{{x}_{k}},\phi_{{x}_{k}}$ of the two scenarios $\zeta^{{{\eta_k}^\star}^{(i)}}_s(t)$ and $ \zeta^{{x}_{k}}_s(t)$, respectively, using \eqref{eq:feature1} and \eqref{eq:feature2}.
		\If {$\mathfrak{d}_f(\zeta^{{{\eta_k}^\star}^{(i)}}_s(t_c), \zeta^{x_k}_s(t_c)) = d(\Delta{\theta}_{k}, \Delta{\phi}_{k})  \leq \frac{\bar{\sigma}_f}{2}$}
		\State $found \gets \textbf{true}$.
		\State \textbf{break}
		\EndIf
		\EndFor
		
		\If{\textbf{not} $found$}
		\State Add the processed individual ${{\eta_k}^\star}^{(i)}$ to the species seed set $X_{{S}_k}$, as it is dissimilar to all existing seeds in $X_{S_k}$, that is,  $X_{S_k} \gets X_{S_k} \cup \{{{\eta_k}^\star}^{(i)}\}$.
		\EndIf
		\EndWhile		
		\State \Return the set of species seeds $X_{{S}_k}$ at the $k$-th generation $G_{k}$.				
	\end{algorithmic}
\end{algorithm}

In Algorithm \ref{alg:identification}, the set of $X_{{S}_k}$ at each generation $G_{k}$ is formed by considering dissimilar individuals $\eta_k^{(i)} \in P_{k}$ in decreasing order of their criticality values, as described in Step 4. The individual with the highest criticality value is denoted by  ${{\eta_k}^\star}^{(i)}$, and is either added as a new species seed if it is sufficiently dissimilar from existing seeds, or discarded otherwise.
To evaluate the dissimilarity, a metric is applied in Step 8 of Algorithm~\ref{alg:identification} to measure the difference between the current candidate the current candidate ${{\eta_k}^\star}^{(i)}$ and each existing species seed $x_{k}$.
In particular, if there is no existing species seed $x_{k} \in X_{{S}_k}$ within half of the dissimilarity threshold $\frac{\sigma_s}{2}$ of the considered individual, then the individual  ${{\eta_k}^\star}^{(i)}$, is added to $X_{{S}_k}$.
To establish a feature-based species seed identification, the dissimilarity metric defined in \eqref{eq:diss} is applied in Step 9 of Algorithm \ref{alg:identification} as follows:
\begin{equation}
	\label{eq:Thf} \mathfrak{d}_f(\zeta^{{{\eta_k}^\star}^{(i)}}_s(t_c), \zeta^{x_k}_s(t_c)) = d(\Delta{\theta}_{k}, \Delta{\phi}_{k}) \leq \frac{\bar{\sigma}_f}{2}, \end{equation}
where
\begin{eqnarray}
	\Delta{\theta}_{k} = \min{(|\theta_{{{\eta_k}^\star}^{(i)}} - \theta_{{x}_{k}}|,|2\pi - (\theta_{{{\eta_k}^\star}^{(i)}} - \theta_{{x}_{k}})| )},	\label{eq:feature1}\\
	\Delta{\phi}_{k}= \min{(|\phi_{{{\phi_i}^\star}_{k}} - \phi_{{x}_{k}}|,|2\pi - (\phi_{{{\eta_k}^\star}^{(i)}} - \phi_{{x}_{k}})|)},
	\label{eq:feature2}
\end{eqnarray}
and the constant $\bar{\sigma}_f$ represents the species distance and can be chosen by the user.
Consequently, species are formed in the system parameter space $\eta$ around a dominant individual ${\eta_k^\star}^{(i)} \in X_{{S}_k}$, referred to as the species seed, by clustering individuals $\eta_k^{(i)} \in P_{k}$ that are similar in the feature space. We highlight that individuals ${\eta_k^\star}^{(i)} \in P_{k}$ whose distance in the feature space is less than $\frac{\bar\sigma_f}{2}$, as determined in Step 9 of Algorithm \ref{alg:identification}, are considered similar to a species seed and are grouped into a species centered around that seed.

After species seeds identification in Algorithm \ref{alg:identification}, a new population of individuals is generated in Steps 6 and 7 of Algorithm \ref{alg:SCGA} by applying standard genetic operators -- selection, crossover, and mutation, as described in \cite{lambora2019genetic}. However, these operations may eliminate some species. Since each species represents a diverse set of individuals in terms of features, a species conservation process, detailed in Algorithm~\ref{alg:feature_conservation}, is employed to ensure their preservation in the next generation $G_{k+1}$.

\begin{algorithm}
	\caption{Feature-Based Species Conservation}\label{alg:feature_conservation}
	\begin{algorithmic}[1]
		\Require{Individual $\eta_{k+1}^{(i)}$ representing the $i$-th system parameter candidate in generation $k+1$; Population at generation $G_{k+1}$: $P_{k+1} := \{ \eta_{k+1}^{(i)}\}_{i=1}^N$; Trajectories $\zeta^{\eta_{k+1}^{(i)}}_s(t)$ and corresponding criticality values $\mathfrak{C}(\zeta^{\eta_{k+1}^{(i)}}_s)$ for all individuals $\eta_{k+1}^{(i)} \in P_{k+1}$, computed in Step 8 of Algorithm~\ref{alg:SCGA}; Set of species seeds $X_{{S}_k}$ from generation $G_{k}$;  Trajectories $\zeta^{x_{k}}_s(t)$ and corresponding criticality values $\mathfrak{C}(\zeta^{x_{k}}_s)$ for all individuals $x_{k} \in X_{S_k}$; Dissimilarity metric $\mathfrak{d}_f$ in \eqref{eq:diss}; Dissimilarity threshold $\bar\sigma_f$.}
		\Ensure{Updated population $ P_{k+1}=\{ \eta_{k+1}^{(i)}\}_{i=1}^N$.}
		\State Mark all individuals $\eta_{k+1}^{(i)} \in P_{k+1}$ at generation $G_{k+1}$ as unprocessed.
		\For{all ${x}_{k} \in X_{S_k}$}
		\State Select the unprocessed individual $\eta_{k+1}^{(i)} \in S'({x}_{k}, \sigma_s)$ with the lowest criticality value,
		where $S'({x}_{k}, \sigma_s)$ is the similarity neighborhood defined in \eqref{eq:species_neighborhood}.
		\If{$\eta_{k+1}^{(i)}$ exists}
		\If{$\mathfrak{C}(\zeta^{\eta_{k+1}^{(i)}}_{s}) < \mathfrak{C}(\zeta^{x_{k}}_{s})$}
		\State Replace $\eta_{k+1}^{(i)}$ with ${x}_{k}$, that is, 
		$\eta_{k+1}^{(i)} \gets x_{k}$.
		\EndIf
		\Else
		\State Select  the unprocessed individual $ \eta_{k+1}^{(i)}\in G_{k+1}$ with the lowest criticality value.
		\State Replace $ \eta_{k+1}^{(i)}\in G_{k+1}$ with ${x}_{k}$, that is, $\eta_{k+1}^{(i)} \gets x_{k}$.
		\EndIf
		\State Mark $\eta_{k+1}^{(i)}$ as processed.
		\EndFor
		\State \Return {Updated population $ P_{k+1}=\{ \eta_{k+1}^{(i)}\}_{i=1}^N$ at generation $G_{k+1}$.}
	\end{algorithmic}
\end{algorithm}

In Algorithm \ref{alg:feature_conservation}, each species seed $x_{k}\in X_{S_k}$ is either conserved or replaced by a better representative of the same species. Therefore, each seed $x_{k}$ with a high criticality value is preserved by replacing a neighboring unprocessed individual ${\eta^{(i)}_{k+1}}$ with a lower criticality value within the following species neighborhood 
\begin{equation}
	\small
	\label{eq:species_neighborhood}
	S'({x}_{k}, \bar{\sigma}_f) \;=\; \{\,{\eta^{(i)}_{k+1}}\in G_{k+1}\mid \mathfrak{d}_f(\zeta^{{x}_{k}}_{s}, \zeta^{{\eta^{(i)}_{k+1}}}_{s})\leq \frac{\bar{\sigma}_f}{2}\}.
\end{equation}

If no such neighbor exists, the seed $x_{k}$ instead replaces the unprocessed individual with the lowest overall criticality value in Step 9 of Algorithm \ref{alg:feature_conservation}. Marking $\eta^{(i)}_{k+1}$ as processed after replacement ensures that the species of each seed $x_{k}$ is preserved in the new generation $G_{k+1}$.

Note that both Steps 5 and 9 of Algorithm~\ref{alg:SCGA} are performed based on Algorithms \ref{alg:identification} and \ref{alg:feature_conservation}, respectively. These procedures follow the same structure as the original SCGA but are adapted to use the feature-based dissimilarity metric \eqref{eq:diss}.

\subsection{A Parameter-Speciation Evolutionary Algorithm}
\label{sec:alg-p}

The parameter-speciation algorithm builds upon traditional SCGA in \cite{li2002species} for scenario-based testing. This method 
follows the same sequence steps as Algorithm \ref{alg:SCGA}, including species seed identification and species conservation. The key differences are the choice of dissimilarity metric and neighborhood definition, which are adapted to the parameter space. In general, this algorithm preserves diversity in the system parameter space, $\eta$,  while its performance in the feature space is not guaranteed. In particular, instead of forming species based on feature-based dissimilarity in Algorithm \ref{alg:identification}, we identify species seeds using the parameter-based dissimilarity metric \eqref{eq:diss2} as follows:

\begin{equation}
	\label{eq:Thp} 
	\mathfrak{d}_p\big(\zeta^{{{\eta_k}^\star}^{(i)}}_s, \zeta^{x_{k}}_s\big) \leq \frac{\bar{\sigma}_p}{2},
\end{equation}
where $\bar{\sigma}_p$ is the species distance constant.
Besides, Algorithm \ref{alg:feature_conservation} is adapted to the parameter space using the following species neighborhood:

\begin{equation}
	\label{eq:Sp}
	S'({x}_{k}, \bar{\sigma}_p) = \Big\{ \eta^{(i)}_{k+1} \in G_{k+1} \ \big| \ \mathfrak{d}_p\big(\zeta^{{x}_k}_{s}, \zeta^{\eta^{(i)}_{k+1}}_{s}\big) \leq \frac{\bar{\sigma}_p}{2} \Big\}.
\end{equation}

\begin{remark}
	We emphasize that Algorithm~\ref{alg:SCGA} improves upon the traditional GA in identifying multiple optima by searching across different species within the population formed based on the similarity among individuals (Step 5) and preserving diversity by copying the seed of each species into the next generation (Step 9). Unlike the parameter-speciation in Section~\ref{sec:alg-p}, which identifies and conserves species based on critiria \eqref{eq:Thp} and \eqref{eq:Sp}, Algorithm~\ref{alg:SCGA} identifies and conserves species by clustering individuals based on dissimilarity in the feature space, as denoted in \eqref{eq:Thf} and \eqref{eq:species_neighborhood}. This use of feature-based clustering increases the probability of conserving individuals that, even if they do not have the highest fitness values, are sufficiently diverse in terms of features compared to the current individuals.
\end{remark}

\begin{remark}
	The species distance thresholds $\bar{\sigma}_f$ and $\bar{\sigma}_p$  affect the performance of SCGA and Algorithm \ref{alg:SCGA} by influencing the formation of species and species seeds. An appropriate value, neither too small nor too large, can be determined through experimental tuning or using estimation methods \cite{li2002species}.
\end{remark}

\subsection{Discussion on  a feature-based diversity measure}
\label{sec:divf}
The performance of the feature-speciation and parameter-speciation approaches is evaluated using the diversity measure in Definition~\ref{def:Q}. In particular, we employ a measure of pairwise dissimilarities between species to identify sets of safety-critical scenarios that are diverse in terms of features. 
Consider the set of $M \in \mathbb{N}$ species seeds at the $k$-th generation defined as 	
\begin{eqnarray}
	\label{eq:seed}
	X_{S_k}=\{x^{(1)}_k,\dots, x^{(M)}_k\}.
\end{eqnarray} 

The feature-based quadratic diversity \eqref{eq:div} of the species seeds \eqref{eq:seed} at the $k$-th generation is computed as 
\begin{equation}
	\label{eq:div2}
	\mathcal{Q}_{k}=\frac{1}{M(M-1)}\sum_{i=1}^{M}\sum_{j=1}^{M} 
	\mathfrak{d}_{f}\!\left(\zeta^{x^{(i)}_k}_{s}(t_c^i),
	\zeta^{x^{(j)}_k}_{s}(t_c^j)\right),
\end{equation} 
where $\zeta^{{x_k}^{(i)}}_{s}(t_c^i)$ and $\zeta^{{x_k}^{(j)}}_{s}(t_c^j)$ are two distinct  scenarios at the most critical moment $t_c^i$ and $t_c^j$.

To compare the performance of the feature-speciation and parameter-speciation approaches, we evaluate the metric at the final generation index $k$. Therefore, a larger value of the metric indicates that the seeds are more widely spread and diverse in the feature space. 

\section{Simulation-based case study}
\label{sec:exa}

\nj{In this section, we apply the proposed optimization framework to generate diverse safety-critical scenarios using the feature-speciation and parameter-speciation algorithms developed in Sections~\ref{sec:alg-f} and~\ref{sec:alg-p}, respectively. We consider two case studies: an intersection environment and a real-world roundabout environment. For each case study, we discuss the results obtained using both approaches. In addition, we evaluate the effectiveness of the feature-speciation method in identifying a diverse set of safety-critical scenarios using the quadratic diversity metric defined in \eqref{eq:div2}.}


\nj{To formulate the optimization problem, we consider the system of interest, $\Sigma$, as either a two-lane intersection or a roundabout} involving two actors, $O_e$ and $O_1$, where $O_e$ denotes the ego vehicle and $O_1$ represents the adversarial vehicle. Besides, we assume two actors are similar in shape---rectangular---and in size in each environment. 

\begin{figure}[t]
	\centering
	\includegraphics[trim= .1in 0 .1in 0, clip, width = 0.44\linewidth]{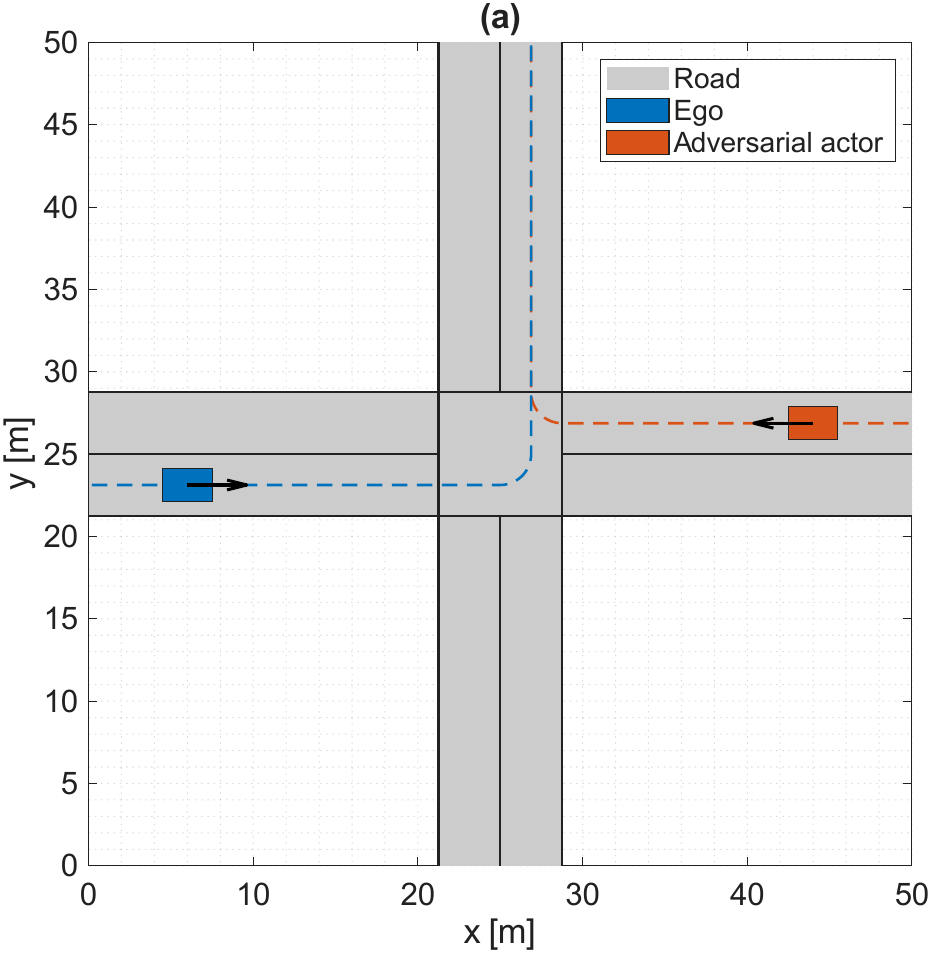}
	 \hspace{0.3cm}
	\includegraphics[trim= .1in 0 .1in 0, clip, width = 0.42\linewidth]{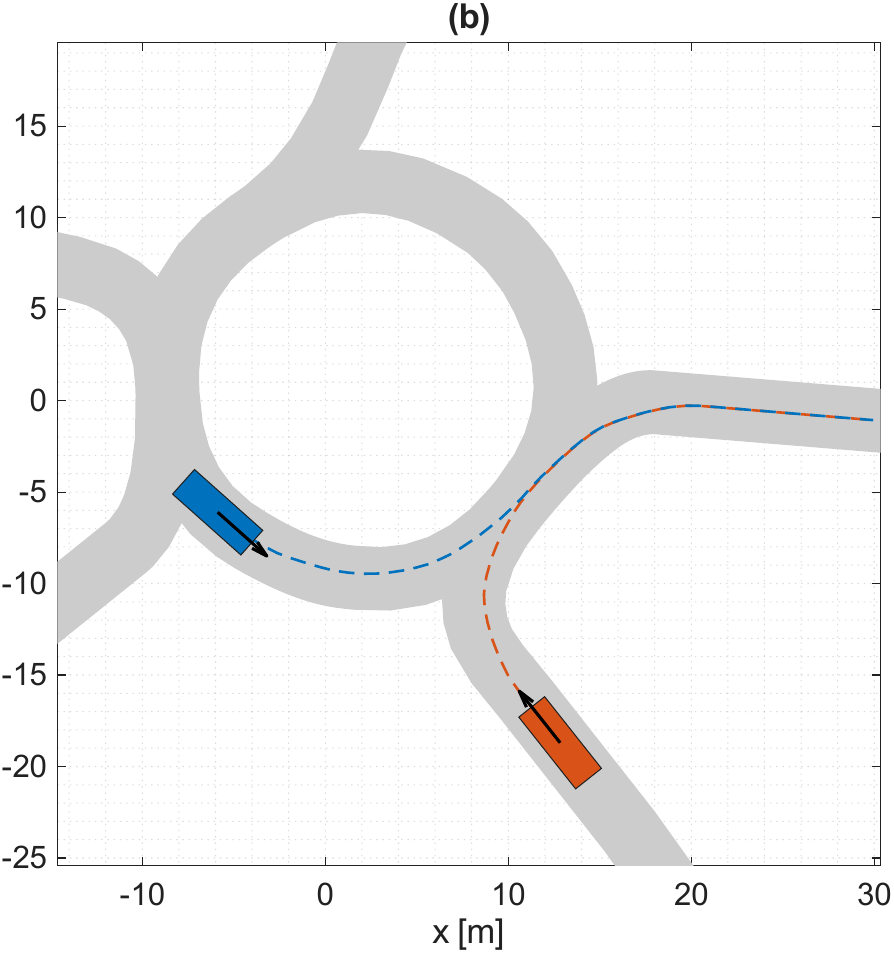}
	\caption{\nj{Intersection environment, denoted by (a), and roundabout environment, denoted by (b), with the ego vehicle $O_e$ in blue and an adversarial vehicle $O_1$ in red.}}
	\label{fig:intersection}
\end{figure}

\nj{The behavioral relations in the behavioral set \eqref{eq:beh} are defined such that both actors follow predefined spatial paths, as illustrated in Fig.~\ref{fig:intersection}(a) and (b). In this figure, the ego vehicle and the adversarial actor approach the intersection and the roundabout, respectively, from opposite directions and take the same exit. The prescribed path in the intersection environment is defined using standardized and smooth geometric motion primitives. The roundabout path is derived from real-world traffic data provided in \cite{althoff2017commonroad}. 
	
To formulate the behavioral set, we denote the prescribed spatial paths by $r_e(s_e(t))$ and $r_1(s_1(t))$, where $s_e(t)$ and $s_1(t)$ are path coordinates of the ego vehicle $O_e$ and the adversarial vehicle $O_1$, respectively.

}

We describe motion and velocities along the path as follows:
\begin{eqnarray}
	\label{eq:path1}
	\dot{s}_e(t)=v_e(t),\\
	\dot{s}_1(t)=v_1(t).
\end{eqnarray}

The Cartesian positions of the actors $O_e$ and $O_1$ are defined by
\begin{eqnarray}
	q_e(t):=r_e(s_e(t)),\qquad
	q_1(t):=r_1(s_1(t)).
\end{eqnarray}

Accordingly, the Cartesian velocities $p_e(t)$  and $p_1(t)$ of the actors $O_e$ and $O_1$, respectively, are obtained as follows
\begin{eqnarray}
	\label{eq:path2}
	p_e(t)=D{q}_e(t)=\frac{d r_e(s_e)}{d s_e}D{s}_e=r_e'\big(s_e(t)\big)\,\dot{s}_e(t),\\	\label{eq:path3}
	p_1(t)=D{q}_1(t)=\frac{d r_1(s_1)}{d s_1}D{s}_1=r_1'\big(s_1(t)\big)\,\dot{s}_1(t).
\end{eqnarray}

The velocity $v_{e}$ along the path  of the ego actor $O_e$ is assumed fixed at $5\ \mathrm{m/s}$.  
The system parameters $\eta$, \nj{defined as the optimization variables in \eqref{eq:OPT}}, are given as follows:
\begin{equation}
	\label{eq:Psys}
	\eta=\begin{bmatrix}
		t_1 ,
		t_e ,
		v_1
	\end{bmatrix}^\top,
\end{equation} 
where  $t_1,t_e \in [0, 10]  \ \mathrm{s}$  are the entry times of the actors $O_1$ and $O_e$ , respectively, into the traffic environment, as depicted in Fig.~\ref{fig:intersection}, and the velocity along the path  of $O_1$ is denoted by $v_1 \in [0, 30] \ \mathrm{m/s}$. Besides, we choose the vectors of VOIs and the latent variables as
\begin{eqnarray}
	\zeta(t)=\begin{bmatrix}
		q_{e}(t)	\\
		q_{1} (t) \\
	\end{bmatrix},\quad
	\alpha=\begin{bmatrix}
		v_e \\
		s_e (t) \\
		s_1 (t) \\
	\end{bmatrix}.
\end{eqnarray} 

The path-following dynamics \eqref{eq:path1}--\eqref{eq:path3} can be written as equality constraint $C_2(\cdot)$ defined in \eqref{eq:beh} as follows

\begin{eqnarray}
	\label{eq:C2}
	\small
	C_2\big(\zeta(t),\eta,\alpha(t),D\alpha(t)\big) \;=\;
	\begin{bmatrix}
		q_e(t) - r_e\big(s_e(t)\big) \\[4pt]
		\dot{s}_e(t) - v_e \\[4pt]
		p_e(t) - r_e'\big(s_e(t)\big)\,\dot{s}_e(t) \\[4pt]
		(s_e(t) - s_{e_0})\,\mathbf{1}_{t_e}(t)\\[4pt]
		q_1(t) - r_1\big(s_1(t)\big) \\[4pt]
		\dot{s}_1(t) - v_1 \\[4pt]
		p_1(t) - r_1'\big(s_1(t)\big)\,\dot{s}_1(t) \\[4pt]
		(s_1(t) - s_{1_0})\,\mathbf{1}_{t_1}(t)\\[4pt]
		v_e - 5
	\end{bmatrix}=\mathbf{0},
\end{eqnarray}
where $s_{e_0}$ and $s_{1_0}$ are considered as the initial conditions and the indicator function is defined as
\begin{eqnarray}
\mathbf{1}_{t_i}(t) =
\begin{cases}
	1, & t = t_i,\\
	0, & t \ne t_i,
\end{cases}
\qquad i \in \{e,1\}.
\end{eqnarray}  

The physical and operational constrains can be written as constraint $C_1(\cdot)$ defined in \eqref{eq:beh} as follows 
\begin{eqnarray}
	\label{eq:C1}
	C_1(\eta) = 
	\begin{bmatrix}
		v_1 - 30 \\[2pt]
		0 - v_1 \\[2pt]
		t_1 - 10 \\[2pt]
		0 - t_1 \\[2pt]
		t_e - 10 \\[2pt]
		0 - t_e
	\end{bmatrix}\preceq \mathbf{0}.
\end{eqnarray}

\nj{Therefore, the behavioral set \eqref{eq:beh}, for the system of interest $\Sigma$, is given as follows 
\begin{equation}
	\label{eq:beh1}
	\mathfrak{B} = \Bigg\{
	\begin{bmatrix}
		q_{e}(t) \\
		q_{1}(t)
	\end{bmatrix}
	\,\Bigg|\,
	\begin{aligned}
		&\eqref{eq:C2},\ \eqref{eq:C1}\text{ hold for all } t\in[0,\mathbb{T}_s],\\
		&\text{with the system parameters } t_1,t_e\in[0,10],\\
		&\text{and } v_1\in[0,30].
	\end{aligned}
	\Bigg\}.
\end{equation}
where $\mathbb{T}_s$ is set to $20\ \mathrm{s}$. Accordingly, any trajectory of $[q_{e}(t),q_{1}(t)]^\top$ belonging to  the behavioral set \eqref{eq:beh1} describes a specific scenario associated with a particular set of system parameters $\eta$ defined in \eqref{eq:Psys}.  
}


The cost function in the optimization problem \eqref{eq:OPT} can be defined as the criticality metric given in \eqref{eq:Critical}, which computes the minimum distance between two actors $O_e$ and $O_1$. In addition, the dissimilarity metric used for clustering population in the optimization algorithm is defined by \eqref{eq:diss}, with equal weights $w_1$ and $w_2$.

\nj{For both case studies, the design parameters for the feature-speciation and parameter-speciation algorithms presented in Sections~\ref{sec:alg-f} and~\ref{sec:alg-p}, respectively, are summarized in Table~\ref{tab:scga_parameters}. To ensure comparability, we use similar parameter values for both algorithms. However, the dissimilarity thresholds are selected differently to account for the differences between the feature and parameter spaces.} 
Due to the stochastic nature of evolutionary-based algorithms, these values are determined experimentally through iterative tuning. The selected parameter values demonstrate appropriate performance in generating diverse scenarios across multiple runs. We perform multiple independent runs, as specified in Table~\ref{tab:scga_parameters}, to ensure statistical reliability rather than relying on a single stochastic trial. In each run, the random seed, initial population, and all stochastic operations of SCGA (selection, crossover, and mutation) are reinitialized. 
\begin{table}[t!]
	\centering
	\caption{Parameters in feature-specification and parameter-specification algorithms in \ref{sec:alg-f} and \ref{sec:alg-p}.}
	\label{tab:scga_parameters}
	\renewcommand{\arraystretch}{1.5} 
	\begin{tabular}{|l|l|}
		\hline
		Population size (N)           & 500 individuals \\ \hline
		Maximum of generations (MaxG) & 60 \\ \hline
		Number of runs                & 10 \\ \hline	
		Dissimilarity threshold ($\bar\sigma_{f}$) & 0.6 \\ \hline
		Dissimilarity threshold ($\bar\sigma_{p}$) & 0.5 \\ \hline
		Crossover rate                 & 1 \\ \hline
		Mutation rate                  & 1 \\ \hline
		Selection function             & Stochastic Universal Sampling \\ \hline 
		$w_1$ and $w_2$                & 0.5 \\ \hline			 
	\end{tabular}
\end{table}
\subsection{Intersection environment}
 \label{sec:Inter} 
\begin{figure}[b!]
	\centering
	\includegraphics[width=0.8\linewidth]{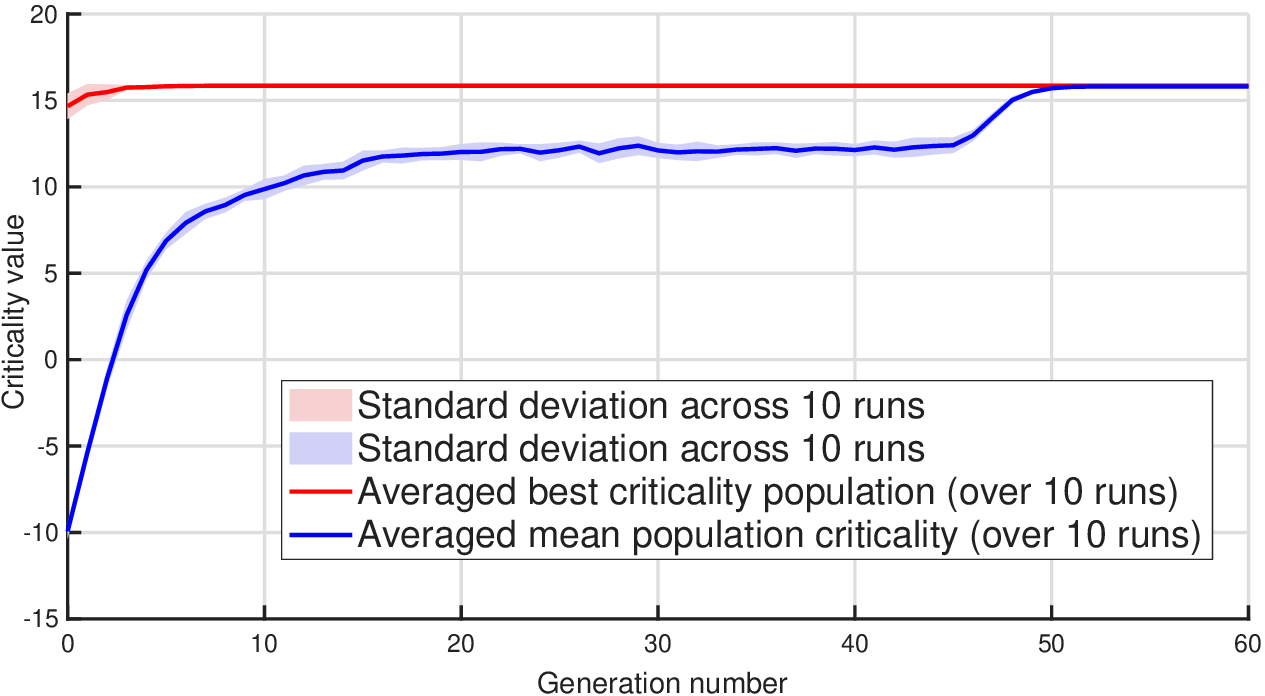}
	\caption{Convergence of population criticality values obtained by the feature-speciation algorithm ~\ref{alg:SCGA}. The solid red curve shows the averaged best criticality (highest individual in the population) across 10 runs, while the solid blue curve represents the mean population criticality averaged over 10 runs. The shaded regions indicate $\pm 1$ standard deviation across the 10 runs. }
	\label{fig:intersection_cost}
\end{figure}

We first focus on intersection case illustrated in Fig.~\ref{fig:intersection}(a). Considering the optimization variables defined in \eqref{eq:Psys}, we apply the feature-speciation evolutionary algorithm described in Section~\ref{sec:alg-f}. Using this algorithm, Fig.~\ref{fig:intersection_cost} illustrates the convergence of the population's best individual (i.e., the one with the highest criticality) and the mean criticality value across all individuals toward the optimal solutions using Algorithm~\ref{alg:SCGA}. The shaded regions in the figure represent the standard deviation across 10 repeated runs of the algorithm, while the solid red and blue curves denote the averaged best and mean population criticalities, respectively. The two curves converge toward each other after approximately 50 generations, indicating that the population has stabilized around the highest criticality value within the population, which may correspond to one or more global optima. Given that the convergence occurs at a positive value, 
the identified safety-critical scenarios correspond to collision cases. 
To analyze the set of diverse safety-critical scenarios generated by Algorithm~\ref{alg:SCGA}, we focus on the results obtained from one representative run--- the 7th run.

\begin{figure}[t]
	\centering
	\includegraphics[width=0.8\linewidth]{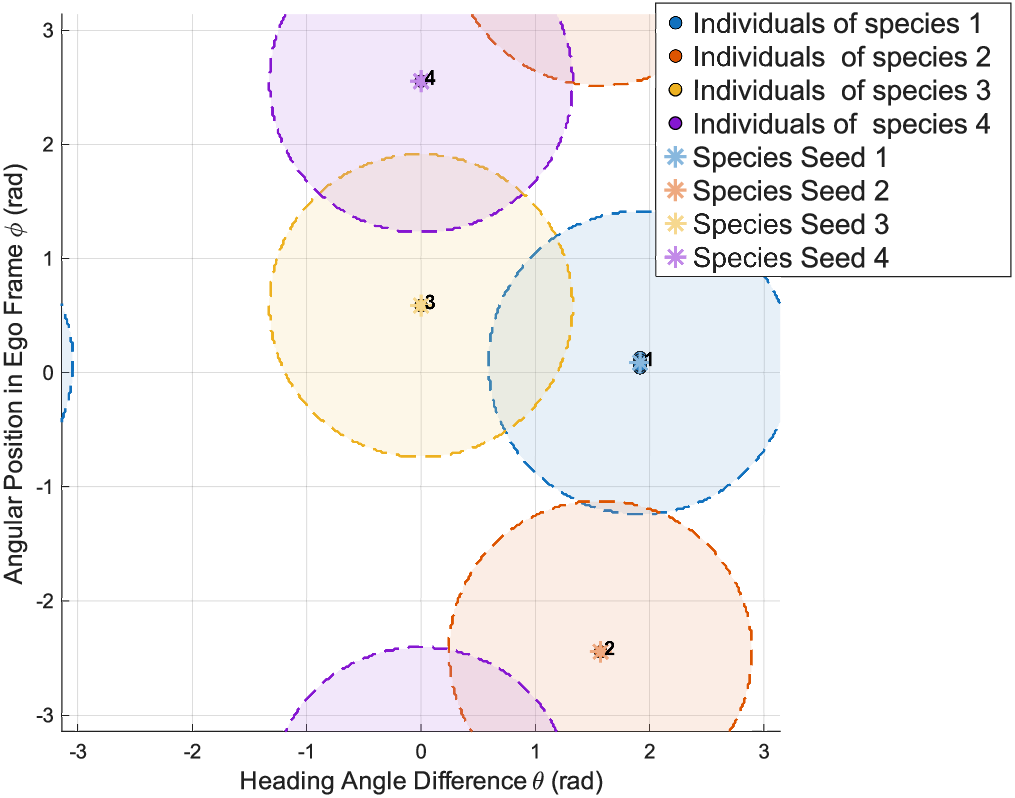}
	\caption{The species seeds obtained by the feature-speciation algorithm \ref{alg:SCGA} in the feature space. The angular position feature $\phi$ and the heading angle feature $\theta$ of all individuals in each species are shown at the final generation $k=60$ for the 7th run in the intersection case.
The species in the feature space are depicted as colorful circles indicating their dissimilarity regions using the feature-based dissimilarity metric \eqref{eq:diss}.
The feature values $\theta$ and $\phi$ of individuals belonging to the same species are shown in the same color, and the feature values of species seeds are marked with stars.
	 }
	\label{fig:intersection_features}
\end{figure}

In Fig.~\ref{fig:intersection_features}, we present the different species formed in the feature space at the final generation of the optimization algorithm \ref{alg:SCGA}, based on the procedures described in algorithms \ref{alg:identification} and \ref{alg:feature_conservation}. 
The final species seeds (or dominant individuals) are given by the set 
\begin{equation}
	\label{eq:seed1}
	X_{S_k} = \{{\eta_k^\star}^{(i_1)}, {\eta_k^\star}^{(i_2)}, {\eta_k^\star}^{(i_3)}, {\eta_k^\star}^{(i_4)}\},
\end{equation}
where ${\eta_k^\star}^{(i_j)} \in P_k$ for $j=\{1,2,3,4\}$ at $k=60$. The positions of these species seeds and other individuals in the final generation within the feature space are shown in Fig.~\ref{fig:intersection_features}. The dissimilarity regions corresponding to each species seed, defined in \eqref{eq:Thf}, are depicted as four colored circles in the same figure. Each dissimilarity region represents the boundary within which individuals possess similar features, i.e., $\theta$ and $\phi$, as the corresponding species seed. Because no two species seeds lie within the same dissimilarity region, it is concluded that each species seed is dissimilar from the others with respect to the dissimilarity metric in \eqref{eq:diss}. Note that each species seed in \eqref{eq:seed1} represents an optimal solution and corresponds to a critical scenario $\zeta^{{\eta_k^\star}^{(i_j)}}_{s}$ computed using the constraints \eqref{eq:C2} and \eqref{eq:C1}.

\begin{figure}
	\centering
	\includegraphics[width=.6\linewidth]{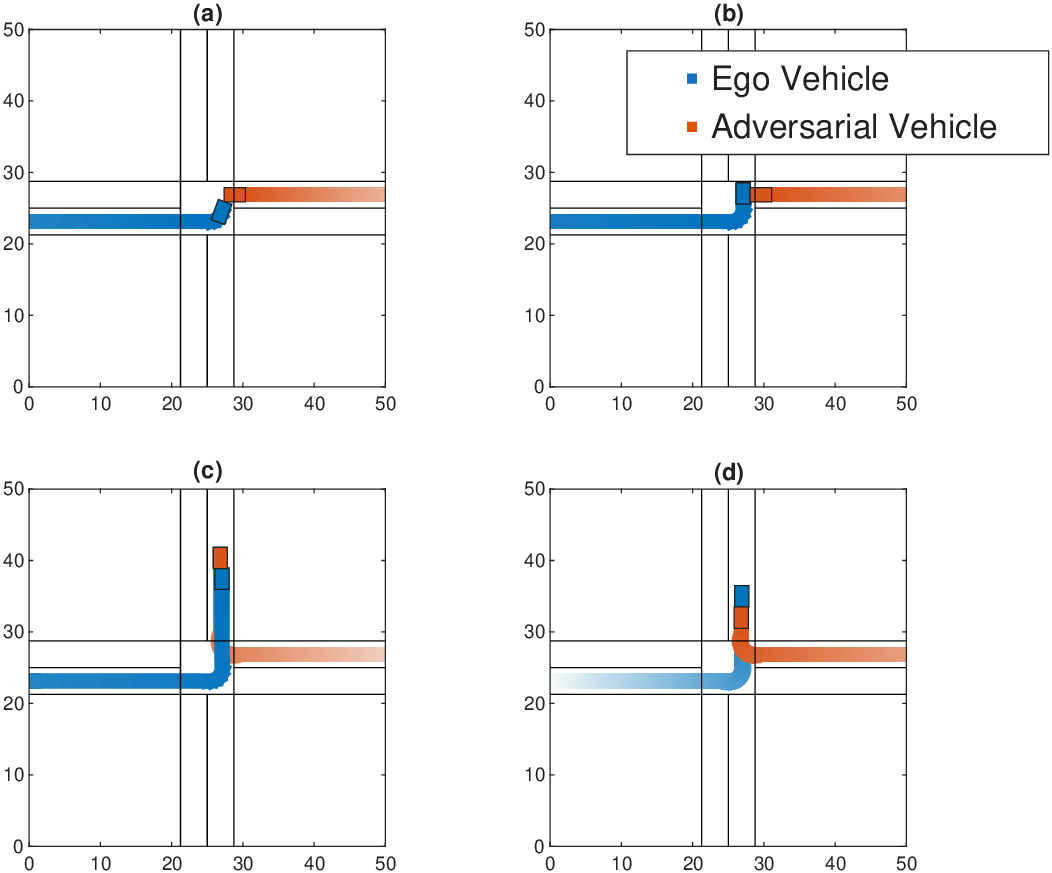}
	\caption{Evaluation of the critical scenarios obtained by the feature-speciation algorithm \ref{alg:SCGA}. The critical scenarios correspond to species seeds in Fig.~\ref{fig:intersection_features} at a given time instant during 7th run of the optimization algorithm in the intersection case. Vehicle positions at the most critical scenes are shown as outlined rectangles, and the paths followed by the vehicles are depicted by shaded trajectory lines, with the ego vehicle $O_e$ shown in blue and the adversarial vehicle $O_1$ in red. }
	\label{fig:intersection_trajectories}
\end{figure}

For better visualization, the critical scenarios corresponding to the species seeds are illustrated in Fig.~\ref{fig:intersection_trajectories}, which shows the trajectories of the ego and adversarial actors up to their most critical scenes. As shown, the safety-critical scenarios correspond to collision cases (i.e., $d_{\min}=0$) and represent dissimilar scenarios characterized by different features $\theta$ and $\phi$ at the most critical scenes. These scenarios include front-end, T-bone, and rear-end collisions. In particular, the safety-critical scenarios (c) and (d) in Fig.~\ref{fig:intersection_trajectories} correspond to front-end collisions with $\theta = 0$, as indicated by species seeds 3 and 4 in Fig.~\ref{fig:intersection_features}. 

\begin{figure}[b!]
	\centering
	\includegraphics[width=0.8\linewidth]{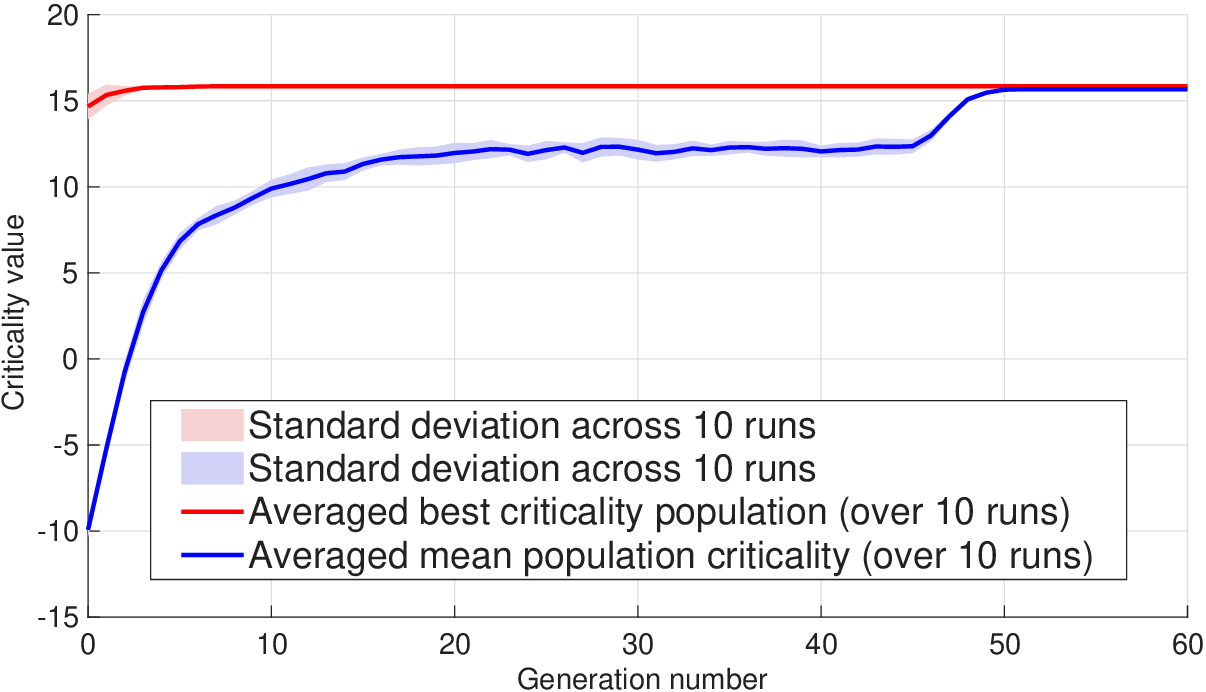}
	\caption{Convergence of population criticality values obtained by the parameter-speciation algorithm in Section \eqref{sec:alg-p}. The solid red curve shows the averaged best criticality (highest individual in the population) across 10 runs, while the solid blue curve represents the mean population criticality averaged over 10 runs. The shaded regions indicate $\pm 1$ standard deviation across the 10 runs.}
	\label{fig:intersection_cost_p}
\end{figure}

As mentioned in Section~\ref{sec:alg-p}, the parameter-speciation algorithm generally maintains diversity in the system parameter space, while its performance in the feature space is not guaranteed. To enable comparison, we apply the parameter-speciation algorithm to generate critical scenarios using the parameter values summarized in Table~\ref{tab:scga_parameters}. The convergence of the population's best individual and the mean criticality values of all individuals over generations is shown in Fig.~\ref{fig:intersection_cost_p}. Similar to the results in Fig.~\ref{fig:intersection_cost}, the two curves converge after approximately 50 generations, indicating that the population has concentrated around the individual(s) with the highest criticality. Consequently, one or more global optima, corresponding to safety-critical scenarios such as collision scenarios, can be considered identified.

\begin{figure}[t]
	\centering
	\includegraphics[width=0.9\linewidth]{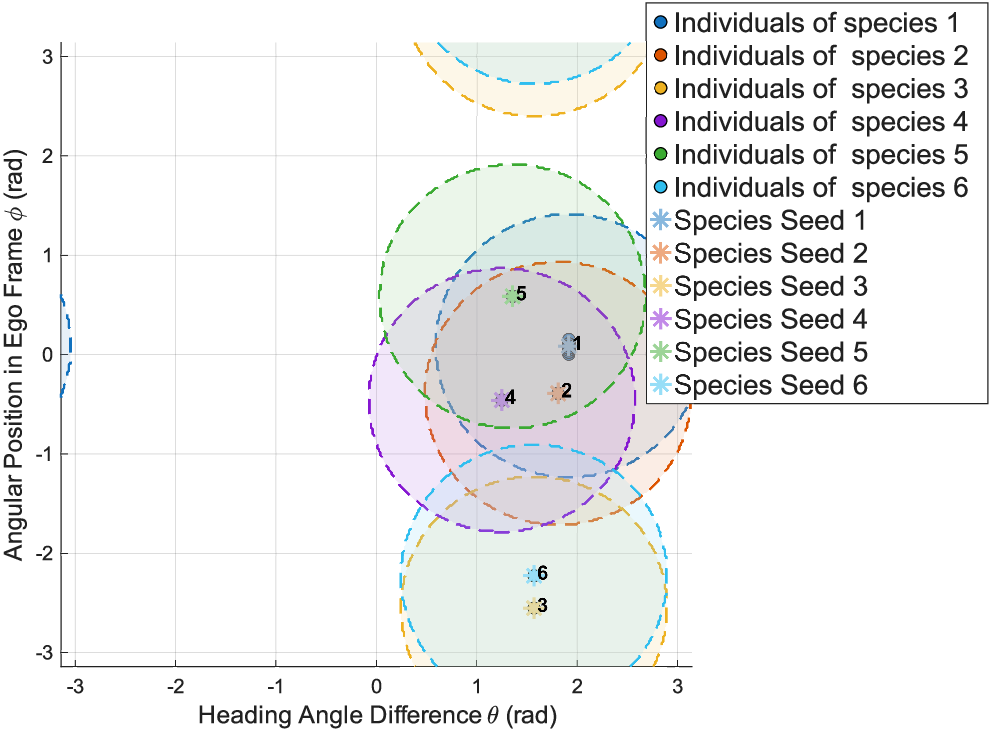}
	\caption{
		The species seeds obtained by the parameter-speciation algorithm in Section \eqref{sec:alg-p} in the feature space. The angular position feature $\phi$ and the heading angle feature $\theta$ of all individuals in each species are shown at the final generation $k=60$ for the 7th run for the intersection case. The species in the feature space are depicted as colorful circles indicating their dissimilarity regions using the feature-based dissimilarity metric \eqref{eq:diss}. The feature values $\theta$ and $\phi$ of individuals belonging to the same species are shown in the same color, and the feature values of species seeds are marked with stars.}
	\label{fig:intersection_features_p}
\end{figure}

Considering the dissimilarity metric \eqref{eq:Thp} with a threshold in  $\bar\sigma_{p}$, we obtain the final species seeds as follows: 

\begin{equation}
	\label{eq:seed2}
	X_{S_k} = \{{\eta_k^\star}^{(q_j)}\}|_{j=\{1,\dots,6\}},
\end{equation}
where ${\eta_k^\star}^{(q_j)} \in P_k$ at $k=60$. 
As explained in Section \ref{sec:alg-p}, the parameter-speciation algorithm identifies species seeds based on the parameter-based dissimilarity metric \eqref{eq:diss2} in the parameter space. Therefore, each species seed in the set \eqref{eq:seed2} is dissimilar from the others in the set \eqref{eq:seed2}  with respect to the system parameters $\eta$ and the parameter-based dissimilarity metric  \eqref{eq:diss2}. To analyze the angle-based features $\theta$ and $\phi$, Fig.~\ref{fig:intersection_features_p} shows each species seeds in the feature space marked with stars. Fig.~\ref{fig:intersection_features_p} illustrates the dissimilarity regions of each species seed based on the feature-based dissimilarity \eqref{eq:Thf}. 
As shown in Fig.~\ref{fig:intersection_features_p}, species seeds 1, 2, 4, and 5, as well as seeds 3 and 6, lie within the same dissimilarity regions, respectively, indicating that they possess similar angle-based features.

\begin{figure}
	\centering
	\includegraphics[width=0.8\linewidth]{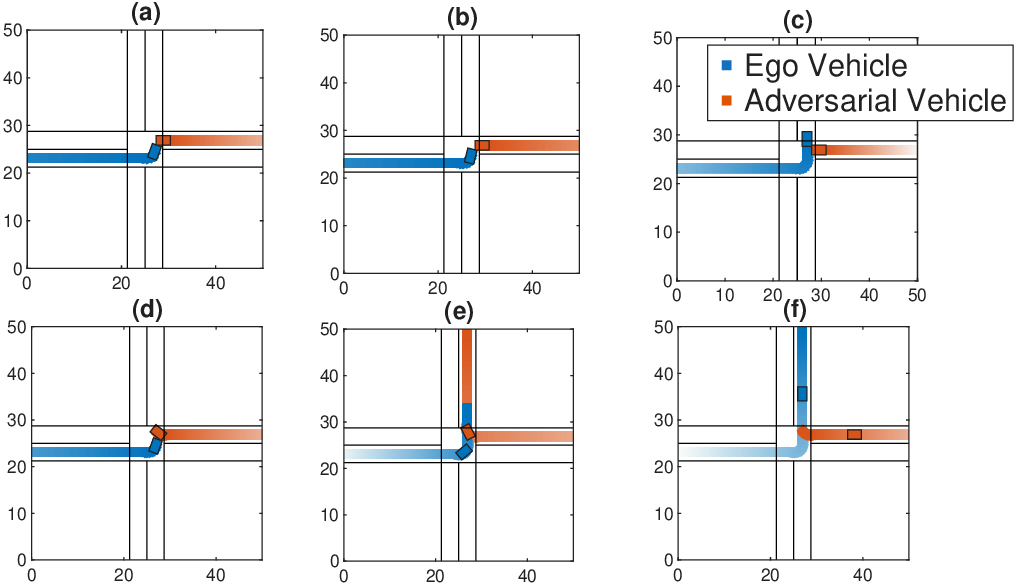}
	\caption{Evaluation of the critical scenarios obtained by the the parameter-speciation algorithm in Section \eqref{sec:alg-p}. The critical scenarios correspond to species seeds in Fig.~\ref{fig:intersection_features_p} at a given time instant during 7th run of the optimization algorithm in the intersection case. Vehicle positions at the most critical scenes are shown as outlined rectangles, and the paths followed by the vehicles are depicted by shaded trajectory lines, with the ego vehicle $O_e$ shown in blue and the adversarial vehicle $O_1$ in red.}
	\label{fig:intersection_trajectories_p}
\end{figure}

The safety-critical scenarios corresponding to the species seeds shown in Fig.~\ref{fig:intersection_features_p} are illustrated in Fig.~\ref{fig:intersection_trajectories_p}. The safety-critical scenarios (a), (b), (d), and (e) belong to similar species in terms of the features $\theta$ and $\phi$, representing T-bone (near-) collision types with front impacts of the ego vehicle $O_{e}$. Besides, the safety-critical scenarios (c) and (f) are similar to each other in terms of the features $\theta$ and $\phi$, corresponding to T-bone near-collision types with possible side impacts of the ego vehicle $O_{e}$.

Comparison of the safety-critical scenarios in Fig.~\ref{fig:intersection_trajectories} and  Fig.~\ref{fig:intersection_trajectories_p} shows the parameter-speciation algorithm generates two groups of dissimilar critical scenarios, i.e., \{(a), (b), (d), (e)\} and  \{(c), (f)\}, while the feature-speciation algorithm identifies four critical scenarios which are dissimilar together in the feature space. In addition, the critical scenarios involving a rear-end collision with the ego vehicle, like (c) and (d) in Fig.~\ref{fig:intersection_trajectories}, are not derived by the parameter-speciation algorithm. Hence, the feature-speciation algorithm \ref{alg:SCGA} generates a set of safety-critical scenarios characterized by dissimilar angle-based features.


\begin{figure}[t]
	\centering
	\includegraphics[width=0.8\linewidth]{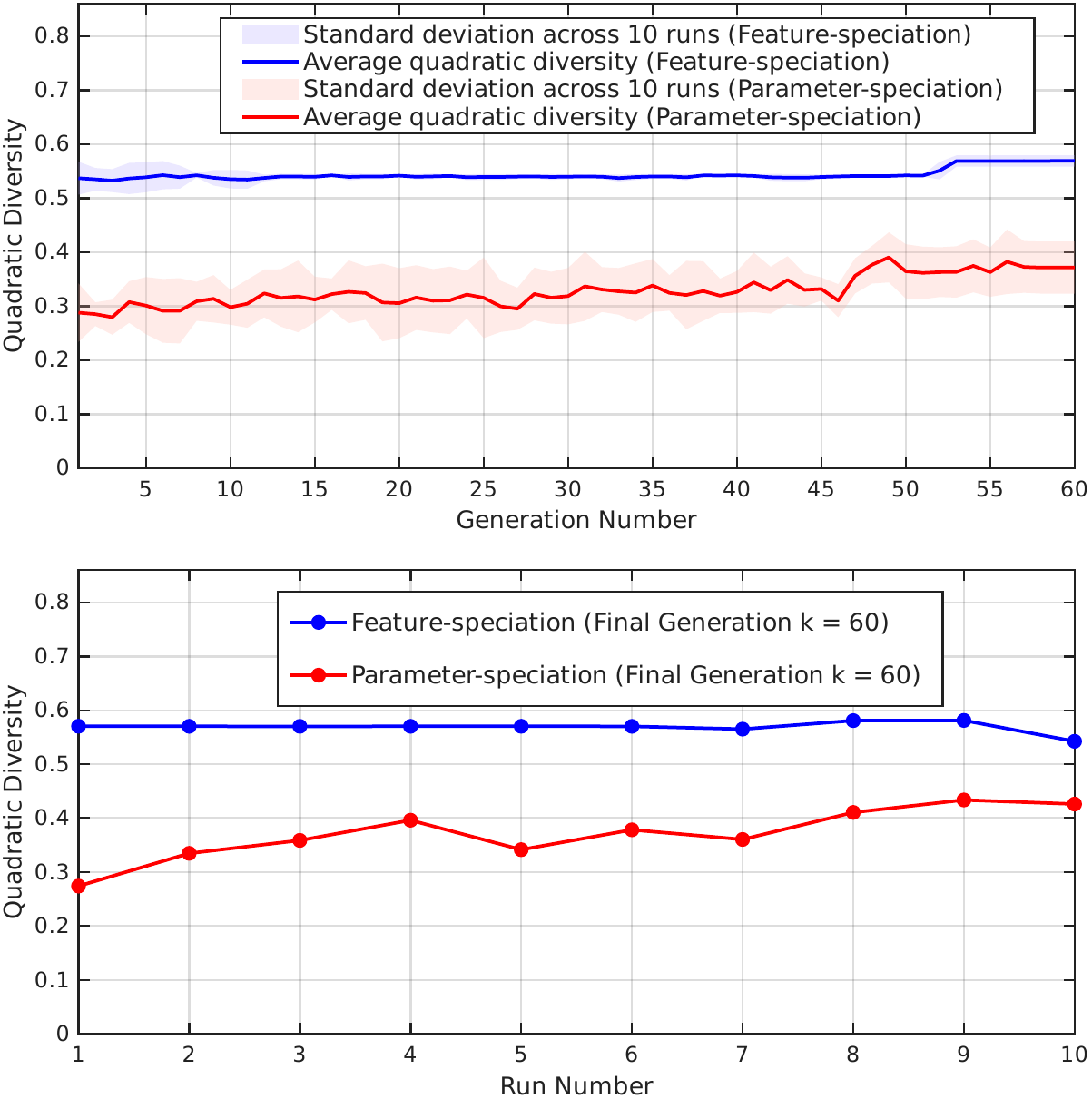}
	\caption{
\nj{Average and final-generation quadratic diversity over 10 runs in the intersection environment. The top plot shows the average quadratic diversity  \eqref{eq:div2} achieved by the feature-speciation algorithm \ref{alg:SCGA} (solid blue curve) and the parameter-speciation algorithm in Section~\ref{sec:alg-p} (solid red curve) over 10 runs. The shaded regions represent ($\pm 1$) standard deviation. The bottom plot shows the quadratic diversity values achieved by the feature-speciation algorithm (blue points) and the parameter-speciation algorithm (red points) at the final generation, (k=60), across the 10 runs.} }
	\label{fig:diversity1}
\end{figure}

To analyze diversity and assess the comprehensiveness of the safety-critical scenarios generated by both approaches, we employ the feature-based diversity measure introduced in \eqref{eq:div2}. In Fig.~\ref{fig:diversity1} and Fig.~\ref{fig:diversity2}, the feature-based quadratic diversity \eqref{eq:div2} is evaluated for both sets of species seeds defined in \eqref{eq:seed1} and \eqref{eq:seed2}. In detail, the mean pairwise feature-based dissimilarity \eqref{eq:diss} among the species seeds defined in \eqref{eq:seed1} and \eqref{eq:seed2} is computed. As shown in Fig.~\ref{fig:diversity1}, the average diversity obtained using the feature-speciation algorithm is higher than that achieved by the parameter-speciation algorithm across generations over 10 runs. A similar pattern is observed in Fig.~\ref{fig:diversity2}, which compares the diversity values at the final generation for both approaches over all runs. Hence, the finial species seeds in \eqref{eq:seed1} are more widely spread compared to \eqref{eq:seed2} in the feature space. Therefore, while the feature-specification algorithm generates four scenarios and the parameter-specification algorithm generates six scenarios in this example, the feature-specification approach still produces greater diversity among the scenarios.
Accordingly, the safety-critical scenario set 
$\hat{\mathcal{E}}_{\text{div}}$
is obtained using \eqref{eq:seed1}, ensuring diversity in the feature space.
\nj{
\subsection{Roundabout environment}

\begin{figure}[b]
	\centering
	\includegraphics[width=0.45\linewidth]{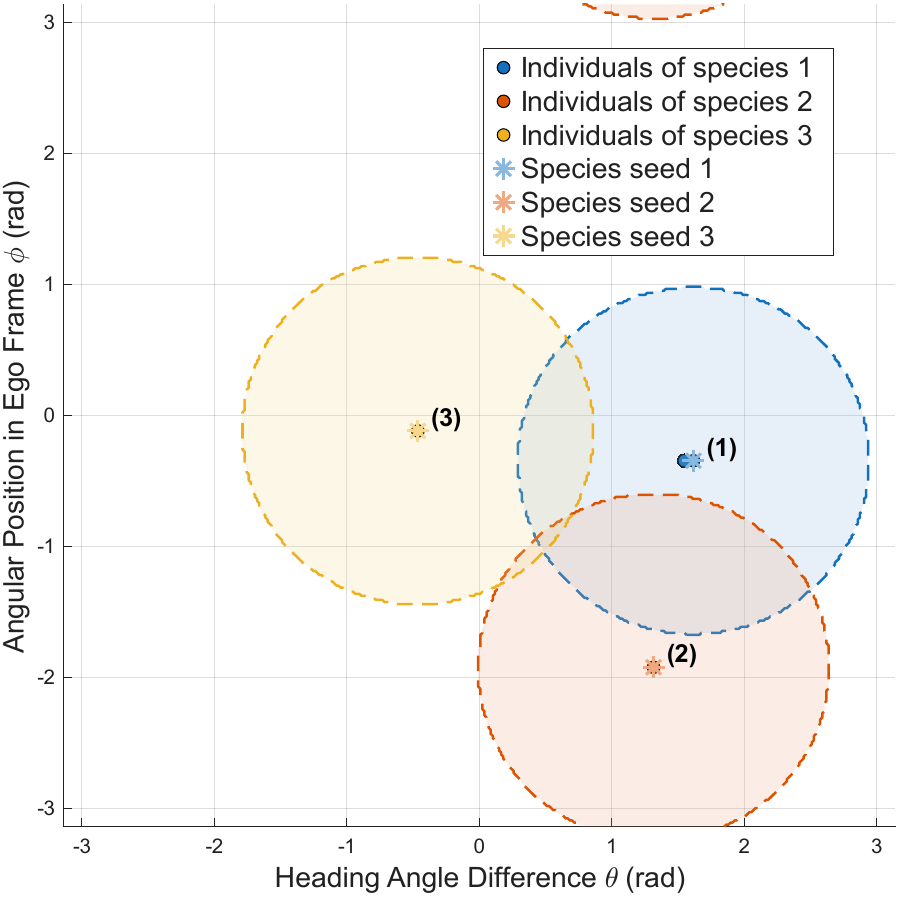}
	\includegraphics[width=0.43\linewidth]{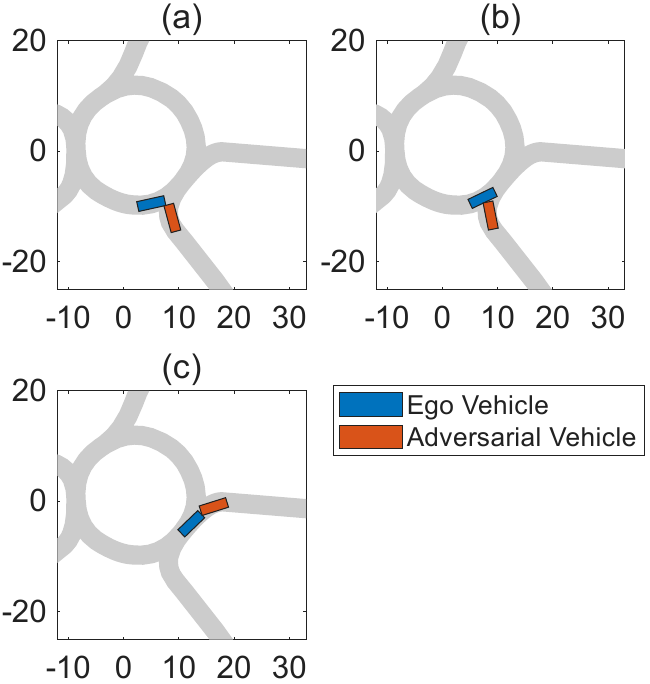}
	\caption{\nj{Feature-space species seeds and corresponding critical scenarios generated by the feature-speciation algorithm \ref{alg:SCGA} in the roundabout environment. The left plot shows the (($\theta,\phi$)) feature values at (k=60) during the 7th run; colors indicate species, circles indicate dissimilarity regions, and stars indicate species seeds. The right plot shows the corresponding critical scenarios, with the ego vehicle ($O_e$) in blue and the adversarial vehicle ($O_1$) in red.}}
	\label{fig:FeatureRound}
\end{figure}
\begin{figure}
	\centering
	\includegraphics[width=0.5\linewidth]{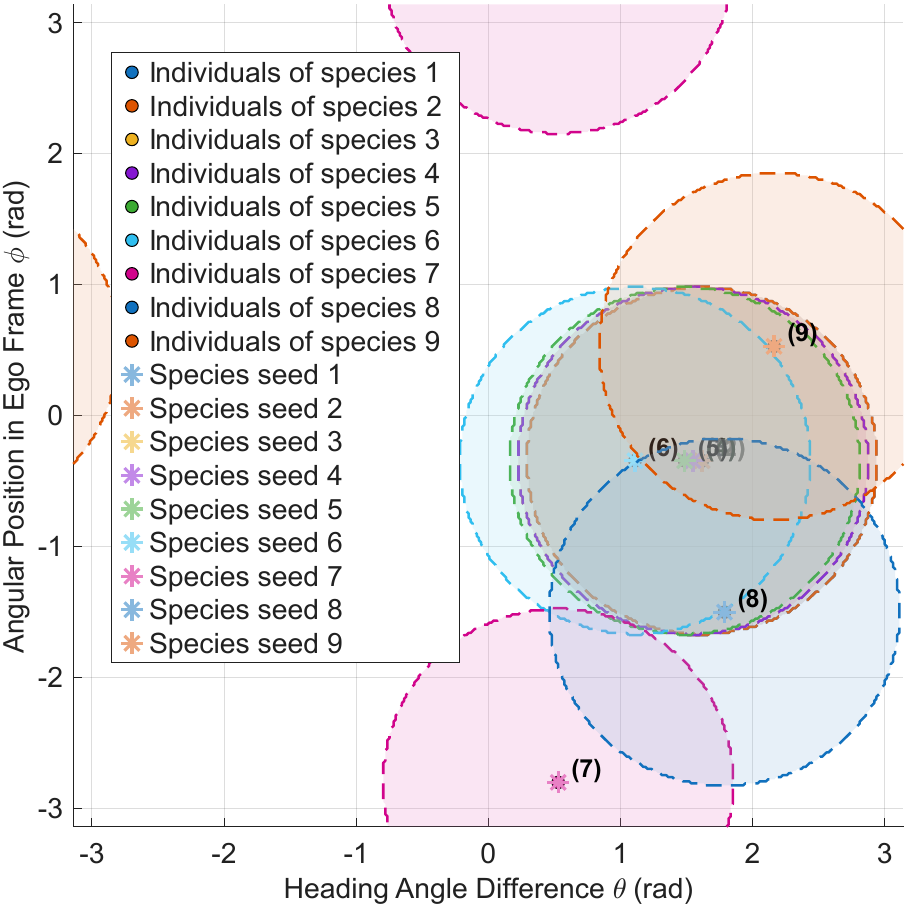}
	\includegraphics[width=0.48\linewidth]{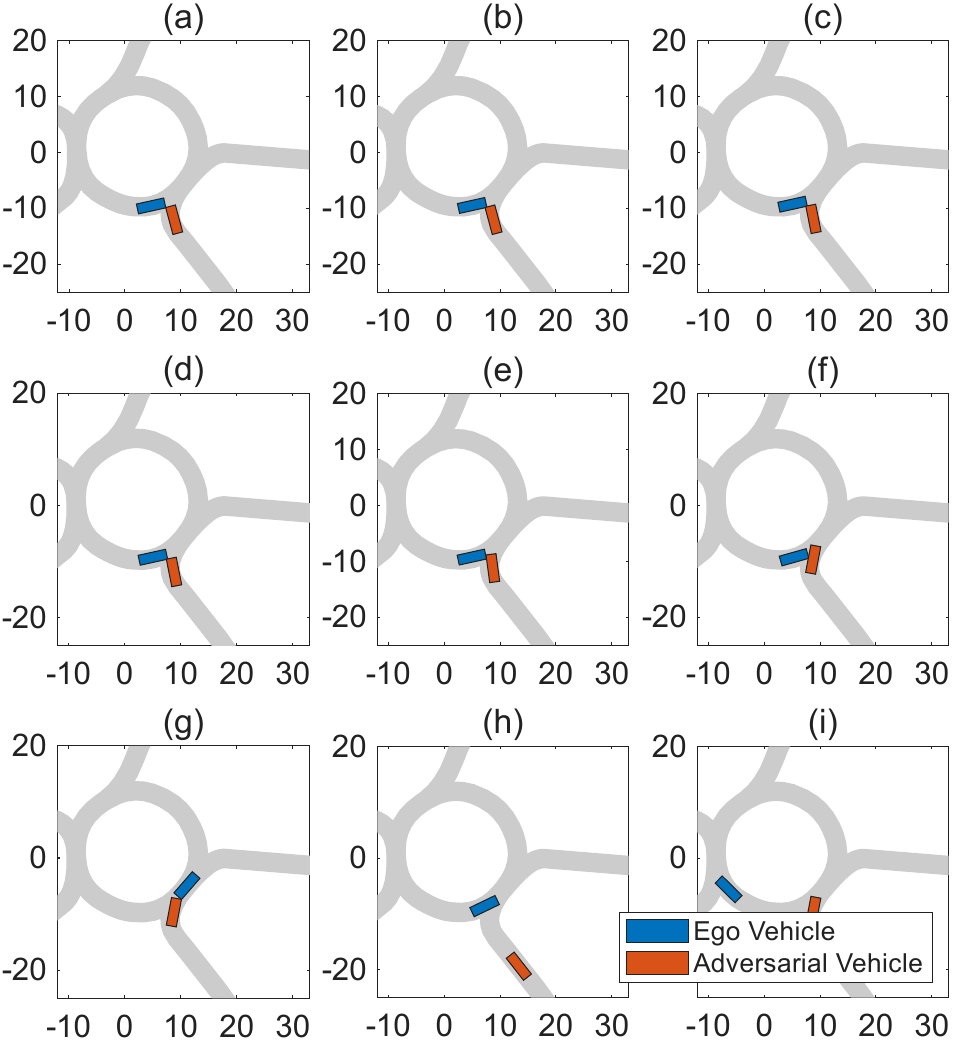}
	\caption{\nj{Feature-space species seeds and corresponding critical scenarios generated by the parameter-speciation algorithm in Section~\ref{sec:alg-p} in the roundabout environment. The left plot shows the (($\theta,\phi$)) feature values at (k=60) during the 7th run; colors indicate species, circles indicate dissimilarity regions, and stars indicate species seeds. The right plot shows the corresponding critical scenarios, with the ego vehicle ($O_e$) in blue and the adversarial vehicle ($O_1$) in red.}}
	\label{fig:ParameterRound}
\end{figure}
\begin{figure}
	\centering
	\includegraphics[width=0.8\linewidth]{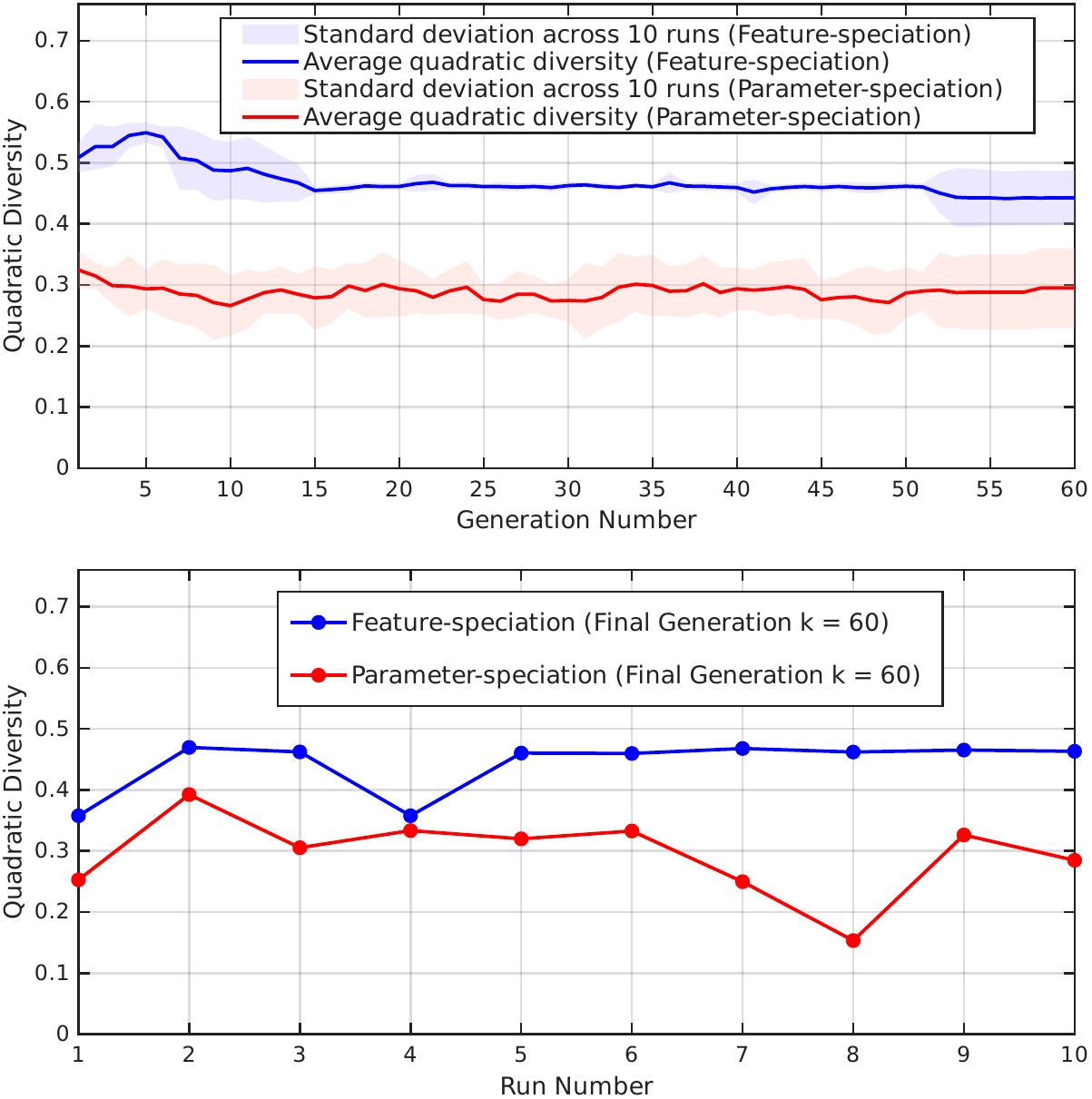}
	\caption{\nj{Average and final-generation quadratic diversity over 10 runs in the roundabout environment. The top plot shows the average quadratic diversity  \eqref{eq:div2} achieved by the feature-speciation algorithm \ref{alg:SCGA} (solid blue curve) and the parameter-speciation algorithm in Section~\ref{sec:alg-p} (solid red curve) over 10 runs. The shaded regions represent ($\pm 1$) standard deviation. The bottom plot shows the quadratic diversity values achieved by the feature-speciation algorithm (blue points) and the parameter-speciation algorithm (red points) at the final generation, (k=60), across the 10 runs.}}
	\label{fig:diversity2}
\end{figure}
To demonstrate the applicability of the proposed method in a real-world setting, we consider a roundabout environment using real-world traffic data from \cite{althoff2017commonroad}. Similar to the intersection case study presented in Section~\ref{sec:Inter}, we use the optimization variables defined in \eqref{eq:Psys}. Using the design parameters summarized in Table~\ref{tab:scga_parameters}, we apply the feature-speciation and parameter-speciation algorithms developed in Sections~\ref{sec:alg-f} and~\ref{sec:alg-p}, respectively. To evaluate the performance of both algorithms, we use the safety-critical scenarios generated during the 7th run as a representative example. 

Following the feature-speciation algorithm, the final species seeds in the feature space are shown in the left plot of Fig.~\ref{fig:FeatureRound}, representing the angle-based features of the generated safety-critical scenarios. The corresponding safety-critical scenarios (a),(b), and (c) for species seeds 1,2, and 3, respectively, denoted by ${\zeta^{\eta_k^\star}_{s}}^{(\lambda_j)} \in P_k$ for $j={1,2,3}$ at $k=60$, are illustrated in the right plot of Fig.~\ref{fig:FeatureRound}. As shown, each species seed lies outside the dissimilarity regions of the other seeds. Therefore, the scenarios shown in Fig.~\ref{fig:FeatureRound} are dissimilar according to the metric defined in \eqref{eq:diss}.

Using the parameter-speciation algorithm, the final species seeds in the feature space, labeled (1)--(9), and their corresponding safety-critical scenarios, labeled (a)--(i), denoted by ${\zeta^{\eta_k^\star}_{s}}^{(\kappa_j)} \in P_k$ for $j={1,\dots,8}$ at $k=60$, are illustrated in Fig.~\ref{fig:ParameterRound}, respectively. As shown in the left plot of Fig.~\ref{fig:ParameterRound}, all species seeds except seed 7 lie within the same dissimilarity region. Therefore, the corresponding generated scenarios yield two dissimilarity categories, \{(a),(b),(c),(d),(e),(f),(h),(i)\} and \{(g)\}, according to the metric defined in \eqref{eq:diss}. Hence, the feature-speciation algorithm identifies a diverse set of three dissimilar safety-critical scenarios, including the Front-to-side collision. 

The feature-based diversity metric in \eqref{def:Q} is used to evaluate the diversity of the safety-critical scenarios generated by both approaches, as shown in Fig.~\ref{fig:diversity2}. As illustrated in the top plot of Fig.~\ref{fig:diversity2}, the average quadratic diversity achieved by the feature-speciation algorithm is higher than the average obtained by the parameter-speciation algorithm across generations over the 10 runs. Besides, the bottom plot shows that the final-generation diversity values are higher for the feature-speciation algorithm across all runs. Therefore, the feature-speciation algorithm generates safety-critical scenarios with greater diversity in the feature space.

}

\section{Conclusion}
\label{se:con}
This work presents an approach for generating diverse safety-critical scenarios for scenario-based testing in automated vehicles. The optimization problem for scenario generation is formulated within a unified mathematical framework based on a behavioral modeling approach. A feature-based species-conserving genetic algorithm is proposed to solve the resulting multimodal optimization problem. The feature-based formulation enables the discovery of multiple critical scenarios that are diverse with respect to angle-based features. This enhances the comprehensive assessment of distinct optima identified in scenario-based testing. The practical applicability of the proposed method is demonstrated through an illustrative case study, confirming its potential for identifying diverse safety-critical scenarios. The approach can also be extended to more complex traffic environments, including highways.

\bibliographystyle{ieeetr}        
\bibliography{references} 

\end{document}